\documentstyle[epsfig]{aipproc}

\begin{document}
\title{Precision Studies of the Higgs Boson\\ 
Profile at the $e^+e^-$ Linear Collider}

\author{Marco Battaglia$^*$ and Klaus Desch$^{**}$ \\
}
\address{$^*$CERN, CH-1211 Geneva 23 Switzerland\\
$^{**}$Universit\"at Hamburg, D-22607 Hamburg Germany}
\maketitle

\vspace*{-0.75cm}
\begin{abstract}
This paper reviews the potential of a high luminosity $e^+e^-$ linear 
collider (LC) in the precision study of the Higgs boson profile.
The complementarity of the linear collider data with that from the LHC 
is also discussed together with a comparison with a muon collider.
\end{abstract}

\section{Introduction \label{sec:intro}}

Explaining the origin of mass is one of the great scientific quests of this
turn of the century. The Standard Model (SM), successfully tested to an 
unprecedented level of accuracy by the LEP and the SLC experiments, addresses 
this question by the Higgs mechanism~\cite{Higgs}. 
The first manifestation of the Higgs mechanism through the Higgs sector is 
represented by the existence of at least one Higgs boson. The observation of a
new spin-0 particle would represent a first sign that the Higgs mechanism of 
mass generation is realised in Nature. This motivates a large experimental 
effort for the Higgs boson search from LEP-2 to the Tevatron and the LHC, 
actively backed-up by new and more accurate theoretical predictions. After a 
Higgs discovery, full validation of the Higgs mechanism can only be 
established by an accurate study of the Higgs boson production and decay 
properties. This paper 
discusses the potential of a high luminosity $e^+e^-$ linear collider (LC) in 
the precision study of the Higgs profile and therefore to the validation of 
the Higgs mechanism of mass generation. In section~II,
the status of Higgs searches, through the LEP-2 program, the forthcoming 
Run-II at the Tevatron and the LHC operation, are shortly discussed and a 
proof of the observability of the Higgs boson at the linear collider, in the 
SM and several of its extensions, also accounting for non-standard couplings, 
is given.
Section~III outlines the landscape of the Higgs production and 
decay properties as it is expected to be depicted by the linear collider data.
These data will tell about the standard or supersymmetric nature of the 
observed Higgs and will allow to determine the supersymmetry parameters in 
the second case. 
Finally, the complementarity of the linear collider data with what will be
learned of the Higgs mechanism in the study of $pp$ collisions at the LHC by 
2010 is discussed and the linear collider potential in Higgs physics is 
compared with that of a muon collider (FMC).

\section{The Quest for the Higgs Boson \label{sec:search}}

The perspectives for the search of the Higgs boson and its detailed study,
depend on its mass $M_H$. In the SM, $M_H$ is expressed as 
$M_H = \sqrt{2 \lambda}v$ where the Higgs field expectation value $v$ is 
determined in the theory as $(\sqrt{2}G_F)^{-1/2} \approx 246$~GeV, while
the Higgs self-coupling $\lambda$ is not specified, leaving the mass as a
free parameter. However, there are strong indications that the mass of the 
Higgs boson in the SM is light. These are derived from the Higgs self-coupling
behaviour at high energies~\cite{triv}, the Higgs field contribution to 
precision electro-weak data~\cite{lepewwg} and the results of direct 
searches at LEP-2 at $\sqrt{s} \ge$ 206~GeV and favour the range
113~GeV $< M_H <$ 206~GeV.

\subsection{From LEP-2 to the LHC}

The LEP-2 $e^+e^-$ collider program has been terminated in November 2000
after reaching centre--of--mass energies up to 209~GeV and having collected
about 850 pb$^{-1}$ above 200 GeV in its last year of operation. Preliminary
results from SM Higgs boson searches ~\cite{pik} have shown an excess of 
events, consistent with Higgs boson production at a mass of approximately 
115 GeV and inconsistent with the expected background at the 2.9$\sigma$ level
when all four LEP experiments are combined. 
While this observation is certainly not firm enough to be considered as 
discovery of the Higgs boson, it represents a first direct hint being 
completely in--line with both the indirect experimental evidence from 
electro--weak radiative corrections~\cite{lepewwg} and the theoretical 
expectations from supersymmetric and grand unified theories.

\begin{figure}[ht!]
\vspace*{-0.75cm}
\begin{center}
\epsfig{file=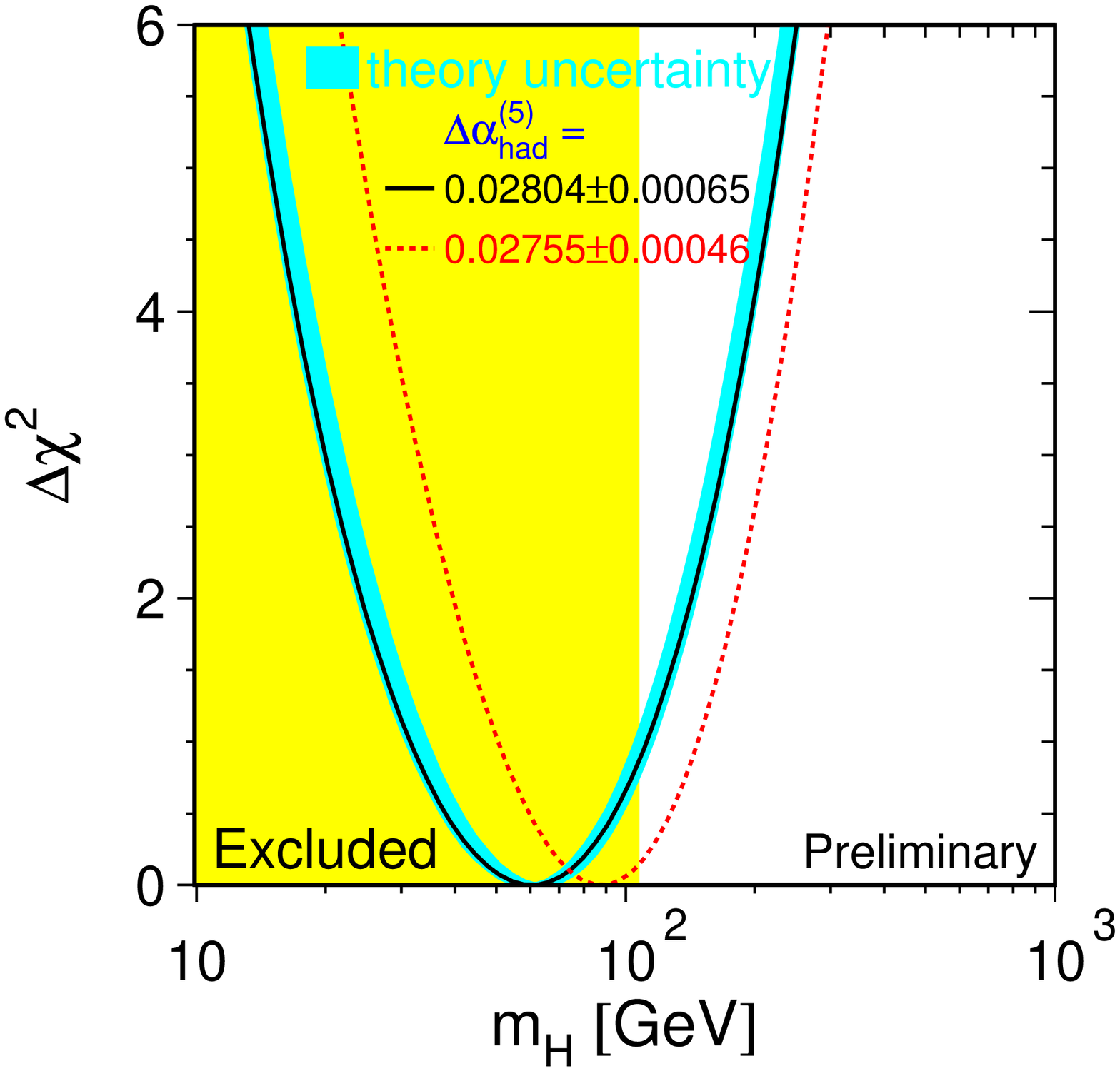,height=6.25cm,width=7.25cm,clip} 
\epsfig{file=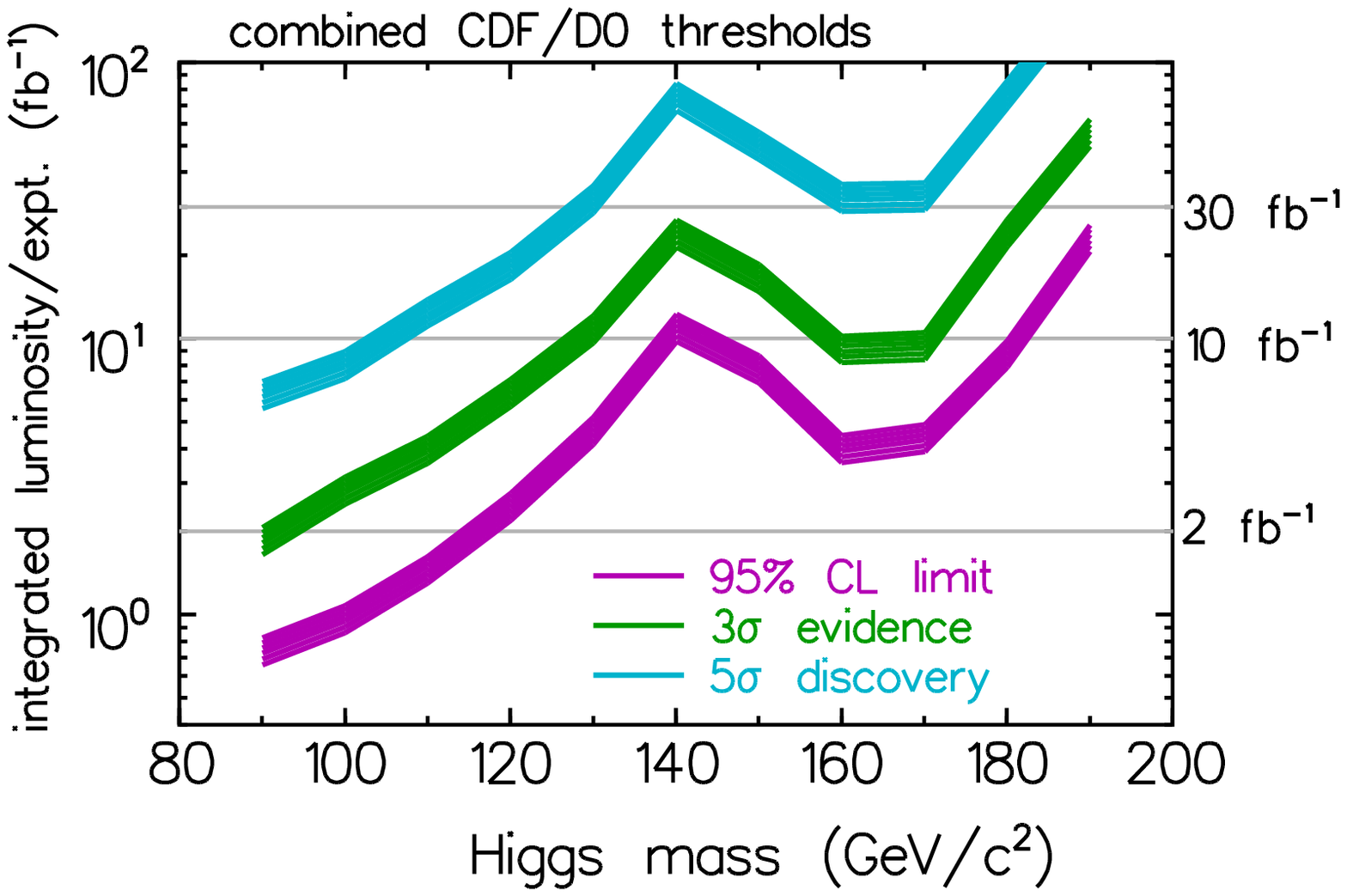,height=6.75cm,width=7.25cm,clip}
\caption{Left: $\chi^2$ of the precision electroweak data as a function of 
the Higgs mass before (thick line) and after (thin line) adding the new BES 
result~[5] in the evaluation of the fine-structure constant $\alpha$ 
(from [4]). 
Right: SM Higgs boson discovery potential of Tevatron Run~II. The integrated 
luminosity is per experiment, assuming both experiments are combined (from
[6]).}
\label{fig:tev}
\end{center}
\end{figure}

After the shutdown of LEP, the experimental search for the Higgs boson 
will continue at the Run~II of the Tevatron $p\bar{p}$--collider. The most
sensitive search channel for $M_H < 130$~GeV is the $W^\pm H$ associated
production with $H\to b\bar{b}$ while a significant part of the sensitivity
at 130~GeV~$< M_H <$~180~GeV comes from the $gg\to H$ process with subsequent
$H\to W^{(*)}W$ decays. The combined sensitivity of the two experiments
CDF and D0 is shown in figure~\ref{fig:tev}. It is interesting to note
that the LEP observation could be ruled out at the 95\% confidence level
already with an integrated luminosity of about 2~pb$^{-1}$ while a $5\sigma$
discovery  would require about ten times more luminosity at the same mass.
A $3\sigma$ evidence could be observed at the Tevatron, in the mass
range up to $M_H =$~180~GeV with 20~fb$^{-1}$.

In $pp$ collisions at $\sqrt{\mathrm{s}}$ = 14~TeV, the LHC will produce a
light Higgs boson mainly through the loop induced $gg$-fusion mechanism and,
with smaller contribution, from the associated productions $WH$ and 
$t\bar{t}H$. The ATLAS and CMS experiments have shown that a Higgs boson, 
with SM couplings, can be discovered over the 
whole theoretically allowed mass range with convincing 
significance~\cite{caner} with an integrated luminosity of 30~fb$^{-1}$, 
through the decay modes $\gamma \gamma$ and $b \bar b$ for 
$M_H <$ 130~GeV and $ZZ^* \rightarrow 4\ell$ for larger masses.
For the MSSM Higgs sector, at least one Higgs boson can be observed for the 
entire $M_A - \tan \beta$ parameter space for an integrated luminosity of
300~fb$^{-1}$.

\subsection{The Linear Collider}

At the LC the Higgs boson can be observed in the Higgs-strahlung production 
process $e^+e^- \rightarrow HZ$ with $Z \rightarrow \ell^+\ell^-$, 
independently of its decay mode, by a distinctive peak in the di-lepton
recoil mass distribution. A data set of 500~fb$^{-1}$ at $\sqrt{s}$ = 350~GeV,
corresponding to one to three years of LC running with the design 
parameters of the different projects proposed, provides a sample of 
3500-2200 Higgs particles produced in the di-lepton $HZ$ channel, for 
$M_H$ = 120-200~GeV. Taking into account the SM backgrounds, dominated by the
$e^+e^- \rightarrow Z^0Z^0$ and $W^+W^-$ productions, the observability of the 
Higgs boson, at the $e^+e^-$ LC, is guaranteed up to its production 
kinematical limit, independently of its decay. At a $\gamma\gamma$ LC, a Higgs
boson with $M_H < 250$~GeV is produced through $\gamma\gamma \rightarrow H$, 
with a cross section approximately an order of magnitude larger than that for 
$e^+e^- \rightarrow HZ$ at the same beam energy. However, the anticipated 
lower luminosity achievable in $\gamma\gamma$ collisions, the large 
backgrounds from $\gamma\gamma \rightarrow {\mathrm{hadrons}}$ and the 
need to analyse predetermined Higgs final states make its detection less 
efficient and model-dependent.  

\section{The Study of the Higgs Boson Profile \label{sec:profile}}

After the observation of a new particle with properties compatible with those
of the Higgs boson, a significant experimental and theoretical effort will be
needed to verify that the observed particle is indeed the boson of the scalar 
field responsible for the electro-weak symmetry breaking and the generation 
of mass. 
Outlining the Higgs boson profile, through the determination of its 
mass, width, quantum numbers, couplings to gauge bosons and fermions and the
reconstruction of the Higgs potential, stands as a very challenging physics
programme. An $e^+e^-$ LC with its large data sets at different
centre-of-mass energies and beam polarisation conditions, the high resolution
detectors providing unprecedented accuracy on the event properties and the
advanced analysis techniques developed from those adopted at LEP and SLC, 
promises to promote Higgs physics into the domain of precision measurements.

\subsection{Higgs Mass}

Since the Higgs mass $M_H$ is not predicted by theory, it is of great 
importance to measure it. Once this mass is fixed, the profile of the Higgs 
particle is uniquely determined in the SM. 
In theories with extra Higgs doublets, such as SUSY, the measurement of the 
masses of the physical boson states is important to predict their 
production and decay properties as a function of the remaining model 
parameters.
At the LC, the Higgs mass can be best measured by exploiting the 
kinematical characteristics of the Higgs-strahlung production process
$e^+e^- \rightarrow Z^* \rightarrow H^0 Z^0$ where the $Z^0$ can be 
reconstructed in both its leptonic and hadronic decay modes.
\begin{figure}[hb!]
\begin{center}
\epsfig{file=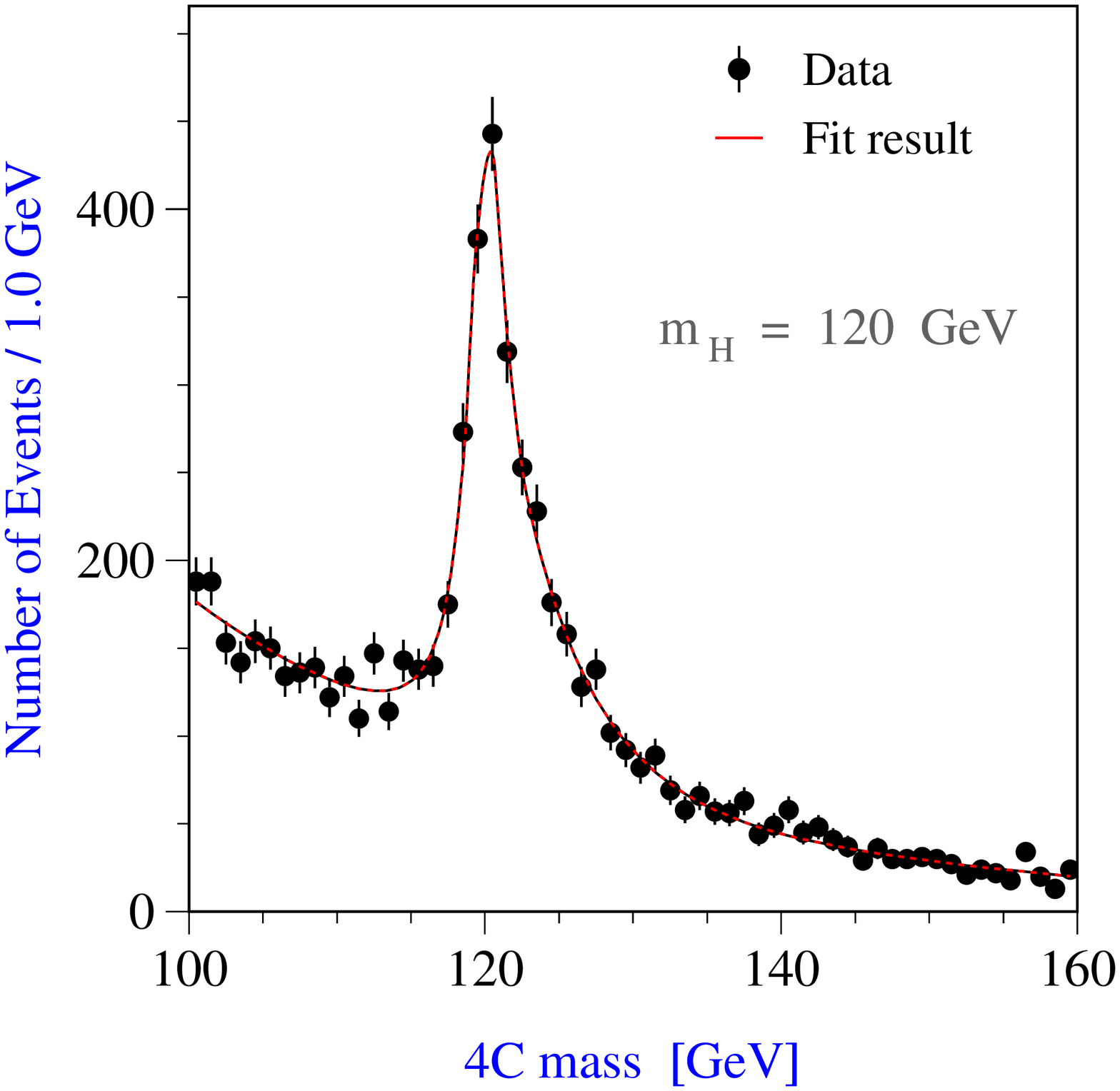,width=6.5cm,height=5.5cm,clip} 
\epsfig{file=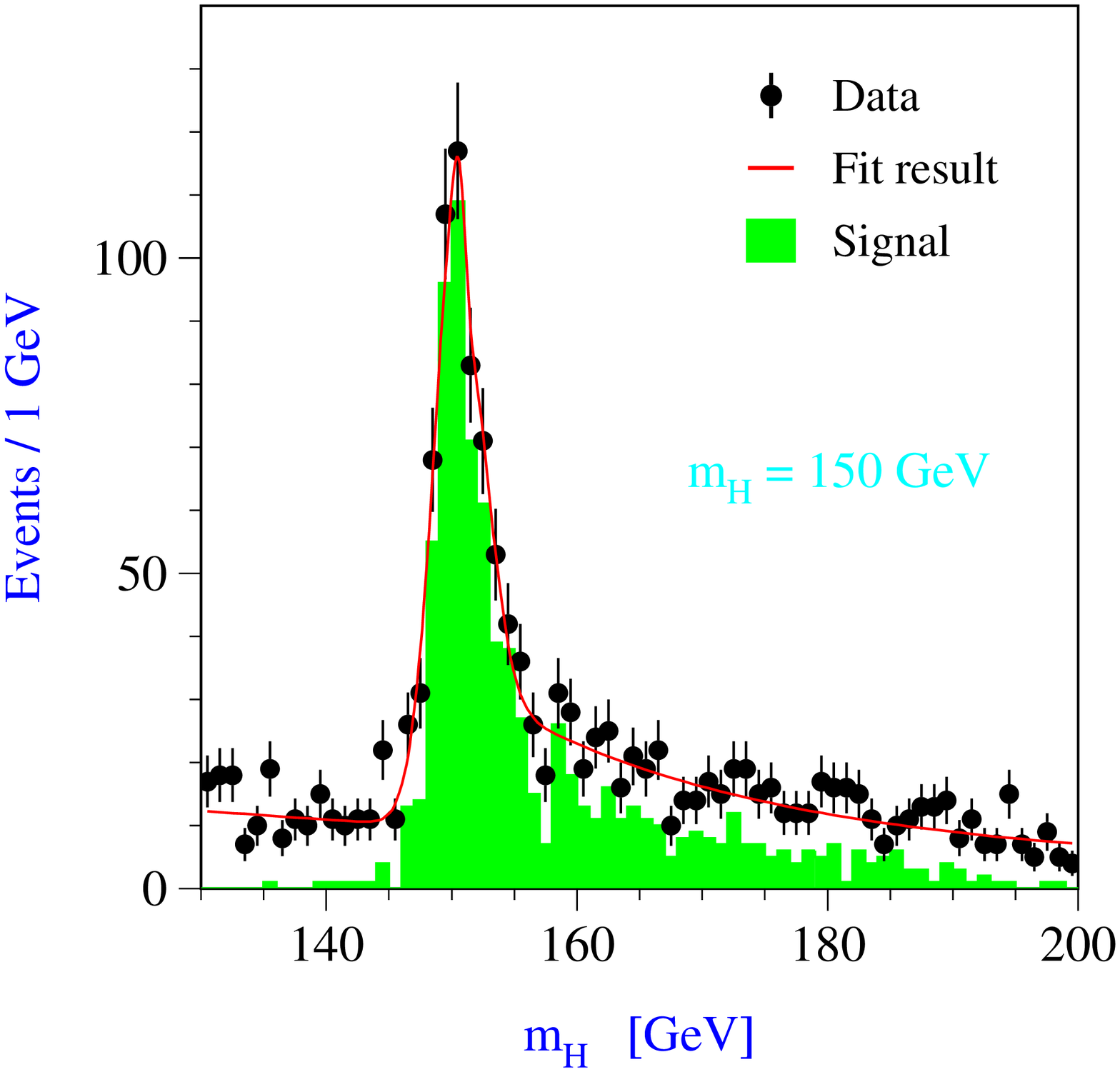,width=6.5cm,height=5.5cm,clip}\\
\vspace{10pt}
\caption{\label{fig:hmass} The Higgs boson mass reconstructed in the 
$H^0 \to b \bar b$, $Z^0 \to \ell^+ \ell^-$ channel for $M_H$=120~GeV (left)
and in the $H^0 \to WW^*$, $Z^0 \to \ell^+ \ell^-$ channel for $M_H$=150~GeV }
\end{center}
\end{figure}
The $\ell^+\ell^-$ recoil mass for leptonic $Z^0$ decays, yields an accuracy of
110~MeV for 500~fb$^{-1}$ of data, without any requirement on the nature
of the Higgs decays. Further improvement can be obtained by explicitly 
selecting $H \rightarrow b \bar b$ ($WW$) for $M_H \le$($>$) 140~GeV.
Here a kinematical 5-C fit, imposing energy and momentum conservation and the 
mass of a jet pair to correspond to $M_Z$, achieves an accuracy of 40 to 
90~MeV for 120$< M_H <$ 180~GeV~\cite{hmass1}.

\subsection{Higgs Quantum Numbers}

The spin, parity, and charge-conjugation quantum numbers $J^{PC}$ of the Higgs
bosons can be determined at the LC in a model-independent way. This allows a 
number of general models, involving $CP$-violating mixture of different Higgs 
bosons, to be tested. The observation of Higgs production at the photon 
collider or of the $H \rightarrow \gamma\gamma$ decay would rule out $J=1$ 
and require $C$ to be positive.
The angular dependence of the $e^+e^-\rightarrow ZH$ 
cross-section allows $J$ and $P$ to be determined and can distinguish the 
SM Higgs boson from a $CP$-odd $0^{-+}$ state $A$, or a $CP$-violating mixture
of the two (generically denoted by $\Phi$ in the following). An additional 
scan of the threshold rise of the Higgs-strahlung cross section can 
unambiguously verify the scalar nature of the observed state~\cite{dmiller}. 
In a general model with two Higgs doublets (2HDM), the three neutral Higgs 
bosons correspond to arbitrary mixtures of $CP$ eigenstates, and their 
production and decay may exhibit $CP$ violation. 
In this case, the amplitude for the Higgs-strahlung process 
can be described by adding a ZZA coupling with strength $\eta$ to the SM 
matrix element. The squared amplitude for the Higgs-strahlung process 
$Z\rightarrow Z \Phi$ is then given by \cite{DK}:
\begin{equation}
|{\cal{M}}|^2 = |{\cal{M}}_{ZH}^{SM}|^2 + \eta 2 Re({\cal{M}}_{ZH}^*
{\cal{M}}_{ZA}) + \eta^2 |{\cal{M}}_{ZA}|^2 \label{matel} 
\end{equation}
\begin{figure}[ht!]
\begin{center}
\begin{tabular} {c c}
\epsfig{figure=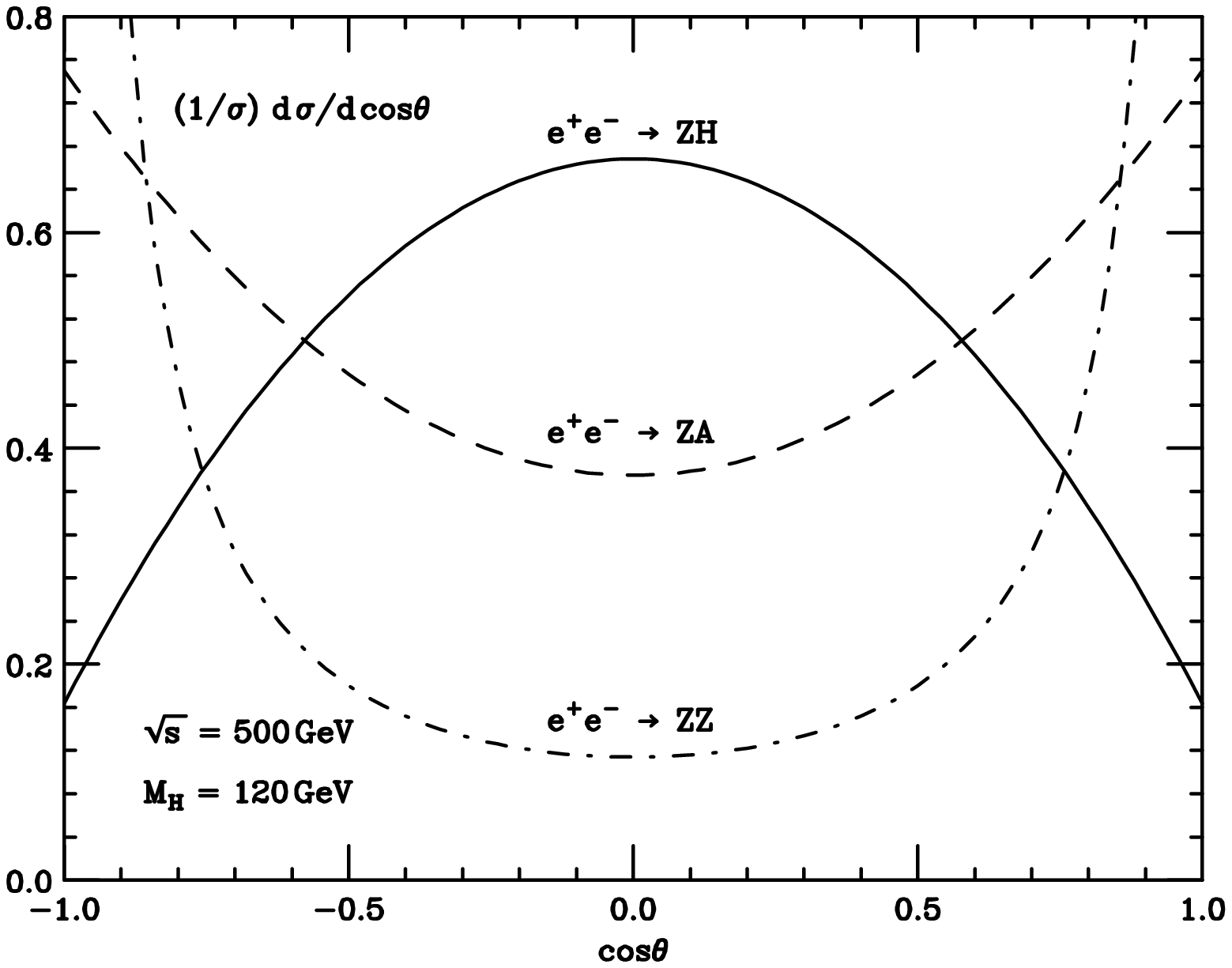,bbllx=80,bblly=210,bburx=550,bbury=580,height=5.5cm,
width=6.25cm} &
\epsfig{figure=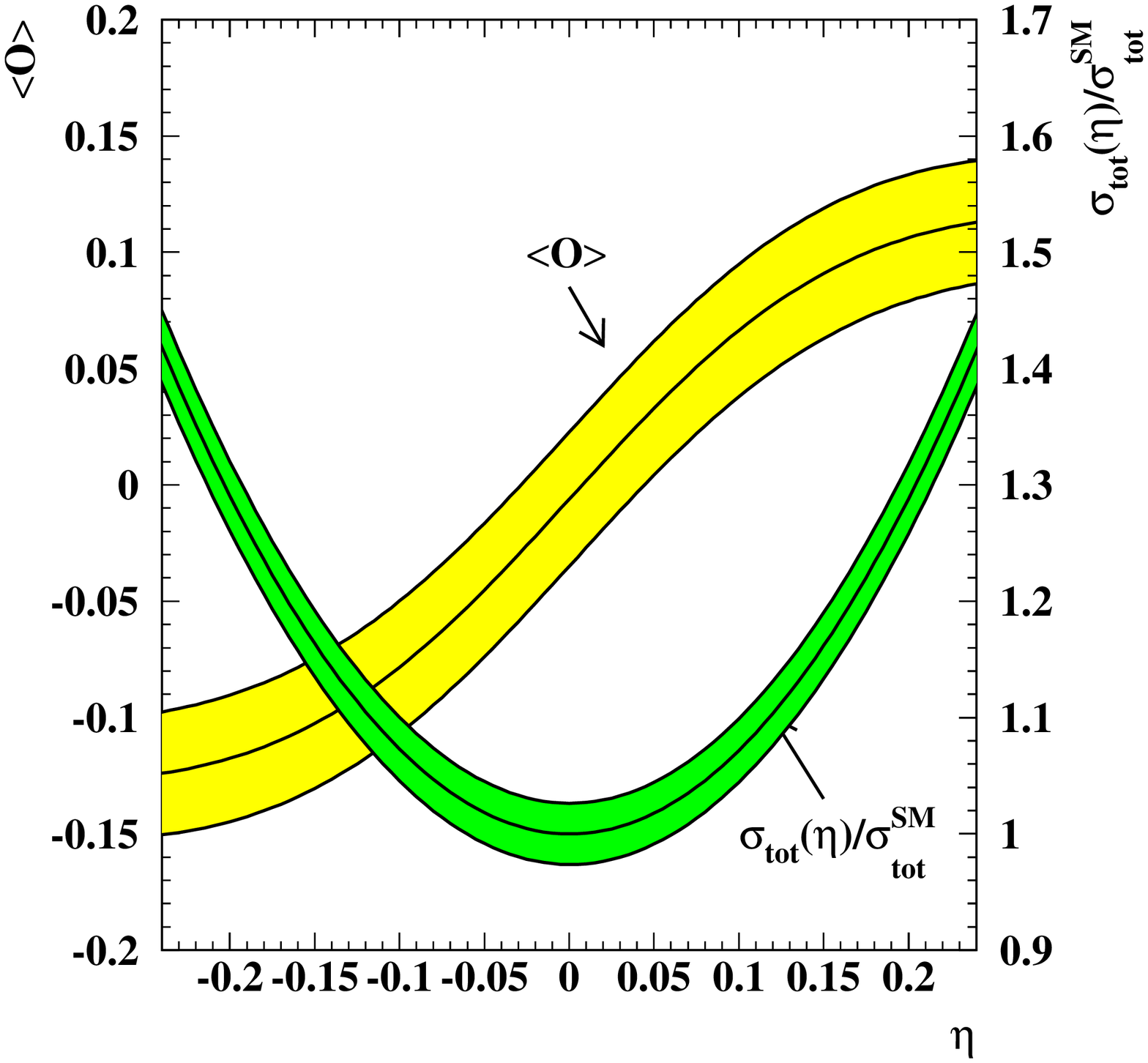,height=6.00cm,width=7.25cm} \\
\end{tabular}
\caption{The $\cos\theta$ dependence of $e^+e^-\to ZH$, $e^+e^-\to ZA$,
$e^+e^-\to ZZ$ for $\protect\sqrt s=500$~GeV, assuming a Higgs boson mass of
120~GeV \protect\cite{BCDKZ} (left) and the dependence of the 
expectation value of the optimal observable and the total cross section
on $\eta$ for $M_H = 120$~GeV,$\sqrt{s}$ = 350 GeV and $\cal{L}$ = 500
$fb^{-1}$ after applying the selection cuts and detector simulation. 
The shaded bands show the $1-\sigma$ uncertainty in the determination of 
$<{\cal{O}}>$ and $\sigma_{Z\Phi}$.}
\label{fig:bak1}
\end{center}
\end{figure}
The first term in $|{\cal{M}}|^2$ corresponds to the SM cross section, the 
second, linear in $\eta$, to the interference term, generates a 
forward-backward asymmetry resulting in a distinctive signal of $CP$ 
violation, while the $CP$-even third term contribution  
$\eta^2 |{\cal{M}}_{ZA}|^2$ increases the total $e^+e^- \rightarrow 
Z\Phi$ cross section.
The angular distributions of the accompanying $Z \rightarrow f \bar f$ decay 
products are also sensitive to the Higgs CP parity and spin as well as to 
anomalous couplings. The information carried by these angular distributions 
has been analysed using the optimal observable formalism for the case of 
500~fb$^{-1}$ of $e^+e^-$ data taken at $\sqrt{s}$ = 350~GeV for a 
120~GeV Higgs boson. The sensitivity to a CP-odd contribution can be 
determined with a sensitivity of better than 3\%~\cite{hqn1}.
In more generality, for the $ZZ\Phi$ coupling 
there may be two more independent $CP$-even terms.
Similarly, there may also be an effective $Z\gamma\Phi$ coupling, generated by
two $CP$-even and one $CP$-odd terms \cite{HIKK} making a total of seven 
complex couplings, $a_Z$, $b_Z$, $c_Z$, $\tilde b_Z$, $b_\gamma$, $c_\gamma$, 
and $\tilde b_\gamma$, where the tilde denotes the $CP$-odd couplings, to be
probed. With sufficiently high luminosity, accurate $\tau$ helicity, good 
$b$ charge identification and polarisation of both beams it will 
be possible to determine these couplings from the angular distribution of 
$e^+e^-\to Z\Phi\to\left(f\bar f\right)\Phi$ as demonstrated by a 
phenomenological analysis~\cite{HIKK}.

\subsection{Higgs Couplings to Fermions}

The SM Higgs couplings to fermion pairs $g_{Hff} = m_f/v$ are fully determined
by the fermion mass $m_f$. The corresponding decay partial widths only depend 
on these couplings and on the Higgs mass. Therefore, their accurate 
determination will represent a comprehensive test of the Higgs mechanism in 
the SM. Further, observing deviations of the measured values from the SM 
predictions probes the parameters of an extended Higgs sector. 
The accuracy of these measurements relies on the performances of the jet 
flavour tagging algorithms and thus on those of the Vertex Tracker, making 
this analysis a major benchmark for optimising the detector design. Several 
analyses have been performed~\cite{brhad0,brhad1,brhad2}. 
The measurement of the decays into $b \bar b$, 
$c \bar c$, $g g$ and $\tau^+ \tau^-$ is based on the selection of Higgs 
decays into two fermions in the $jjjj$, $jj \ell \ell$ and $jj + E_{miss}$ 
topologies. The decay rates for the individual hadronic modes are extracted by
a likelihood fit to the jet flavour tagging response, while the $\tau$ final 
states are selected by a dedicated likelihood, based on vertexing and 
calorimetric response to separate the $H \to \tau^+\tau^-$ from the hadronic 
decays. For $M_H \le$ 140~GeV, the hadronic modes have branching fractions 
that are large enough in the SM to be measured to an accuracy better or 
comparable to their theoretical uncertainties. For larger values of the Higgs 
boson mass, as the $WW^*$ decay becomes predominant, the $H \to b \bar b$ 
decay can still be measured with an accuracy better than 10\% up to 170~GeV. 
\begin{figure}[htb]
\begin{center}
\epsfig{file=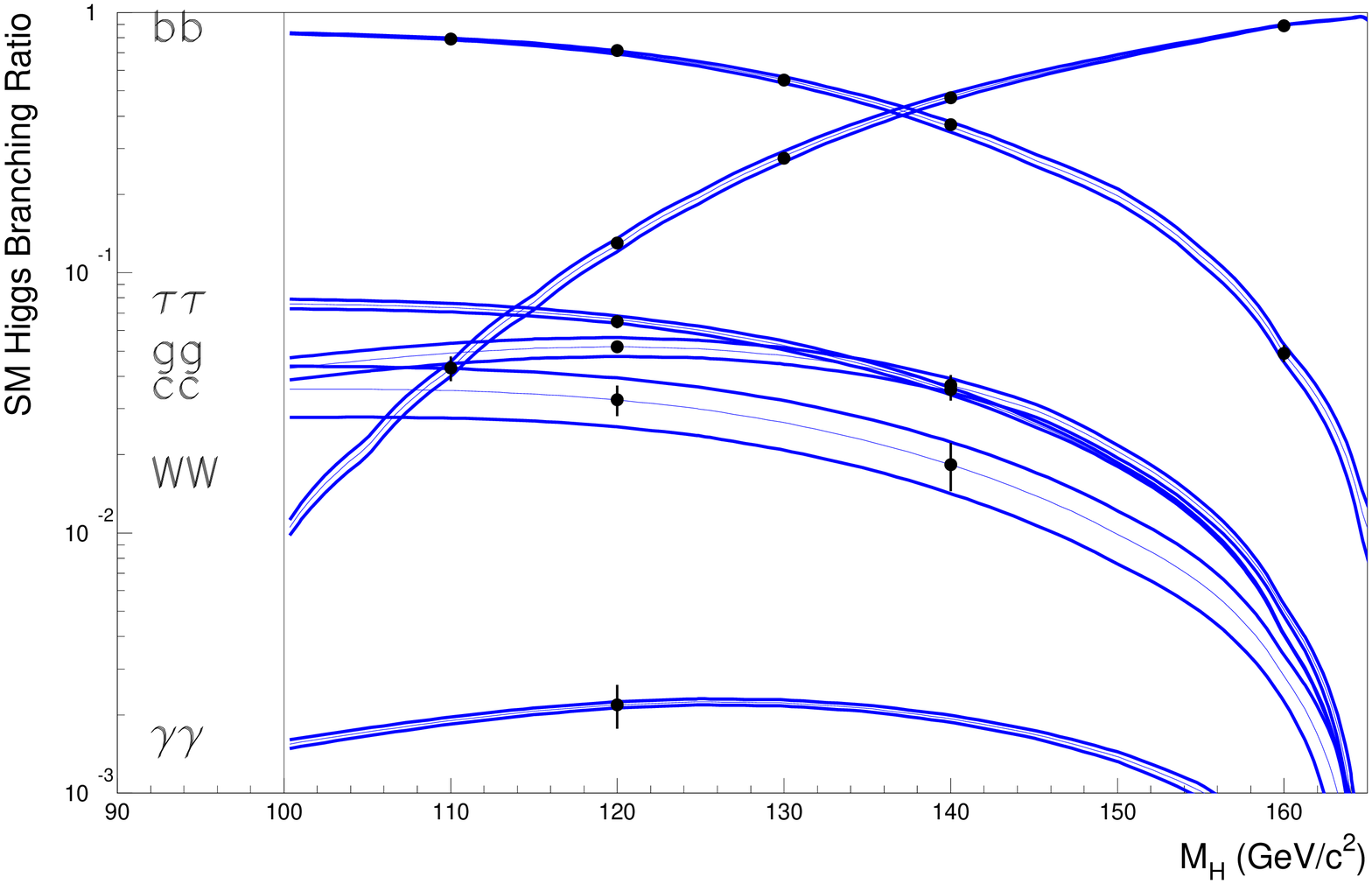,width=11.0cm,clip}
\caption{SM predictions for the Higgs boson decay branching ratios as a 
function of $M_H$. Points with error bars show the expected experimental 
accuracy, while the lines show the estimated uncertainties on the SM 
predictions due to the value of the fermion masses and of $\alpha_s$.} 
\end{center}
\label{fig:hbr}
\end{figure}

\begin{table}[hb!]
\begin{center}
\caption{Relative accuracy in the determination of Higgs 
boson decay branching ratios for 500~fb$^{-1}$ at $\sqrt{s}$ = 350~GeV
\label{table:brsummary}}
\vspace{0.4cm}
\begin{tabular}{|c|c|c|c|}
\hline
Channel & $M_H$ = 120 GeV & $M_H$ = 140 GeV & $M_H$ = 160 GeV \\
\hline \hline
$H^0 / h^0 \rightarrow b \bar b$        & $\pm$ 0.024 & $\pm$ 0.026 & 
$\pm$ 0.065\\
$H^0 / h^0 \rightarrow c \bar c$        & $\pm$ 0.085 & $\pm$ 0.190 &   \\
$H^0 / h^0 \rightarrow g g$             & $\pm$ 0.055 & $\pm$ 0.140 &   \\
$H^0 / h^0 \rightarrow \tau^+ \tau^-$   & $\pm$ 0.050 & $\pm$ 0.080 &   \\
\hline
\end{tabular}
\end{center}
\end{table}

The Higgs coupling to the top quark, is the largest coupling in 
the SM ($g_{Htt}^2 \simeq 0.5$ to be compared with $g_{Hbb}^2 
\simeq 4 \times 10^{-4}$). However, for a light Higgs boson this coupling is 
accessible indirectly in the loop process $H \rightarrow g g$ and directly 
only in the Yukawa process $e^+e^- \rightarrow t \bar t H$~\cite{ttH}. 
This process has a cross section of the order of only 0.5~fb for 
$M_H \sim 120$ GeV at $\sqrt{s}$ = 500~GeV and 2.5~fb at $\sqrt{s}$ = 800~GeV.
The QCD corrections have been calculated recently up to 
next-to-leading order~\cite{ttHth} and were found to be large and positive at 
$\sqrt{s} \sim 500$~GeV because of threshold effects, while they become small 
and negative at $\sqrt{s} \sim 1$~TeV. The distinctive signature, consisting 
of two $W$ bosons and four $b$-quark jets, makes it possible to isolate these 
events from the thousand times larger backgrounds. 
In consideration of the small statistics, the analysis uses a set of 
highly efficient pre-selection criteria and a Neural Network trained to 
separate the signal from the remaining backgrounds. Because the backgrounds 
are so large, it is crucial that they should be well modelled both in absolute
level and in the event shapes which determine how they are treated by the 
Neural Net. For an integrated luminosity of 1000~fb$^{-1}$ at $\sqrt{s}$ = 
800~GeV, the uncertainty in the Higgs-top Yukawa coupling after combining the 
semileptonic and the hadronic channels is $\pm$ 4.2\%~(stat), for $M_H$ = 
120~GeV, becoming $\pm$ 5.5\%~(stat+syst) by assuming a 5\% uncertainty in the
overall background normalisation~\cite{tthlight}. 

If the Higgs boson mass is larger than the kinematical threshold for $t \bar t$
production, the Higgs Top Yukawa couplings can be measured from the 
$H \rightarrow t \bar t$ branching fractions, similarly to those of the other
fermions discussed previously in this section. A study has been performed based
on the analysis of the process $e^+e^- \rightarrow \nu_e \bar \nu_e H 
\rightarrow \nu_e \bar \nu_e t \bar t$ for 350~GeV$<M_H<$ 500~GeV
at $\sqrt{s}$ = 800~GeV. The $e^+e^- \rightarrow t \bar t$ and the 
$e^+e^- \rightarrow e^+e^- t \bar t$ backgrounds are reduced by an event 
selection based on the characteristic event signature with six jets, two of 
them from a $b$ quark, on the missing energy and the mass.
Since the S/B ratio is expected to be large,
the uncertainty on the top Yukawa coupling is dominated by the statistics and 
corresponds to 7\% (15\%) for $M_H$ = 400 (500) ~GeV for an integrated
luminosity of 500~fb$^{-1}$~\cite{ttheavy}.

\subsection{Higgs Couplings to Massive Gauge Bosons}

In the SM the coupling of the Higgs boson to the massive gauge bosons is
given by $g_{HVV} = 2 M_V^2 / v$ for $V=W,Z$. The ratio
of the $W^\pm$ and $Z^0$ couplings is dictated by the $SU(2)\times U(1)$ 
symmetry and thus valid in any model obeying this experimentally well
established symmetry. 

At the LC, both couplings can be probed independently with high accuracy. 
The $g_{HZZ}$ coupling is most sensitively probed through the measurement
of the cross section for the Higgs--strahlung process, $e^+e^-\to H^0Z^0$,
which at tree level is proportional to $g_{HZZ}^2$. Since the recoil mass
method allows to extract this cross section independently of the subsequent
Higgs boson decay, no further model assumptions need to be made. Detailed
experimental studies have shown that the Higgs--strahlung cross section 
can be measured with accuracies between 2.4\% and 3.0\% for Higgs boson 
masses between 120 and 160~GeV~\cite{hmass1} 
only deteriorating slowly for higher Higgs boson masses due to the 
decreasing production cross section.

The $g_{HWW}$ coupling is probed both through the measurement of the 
cross section for the $WW$--fusion process and the decay branching ratio
for $H^0\to WW^*$. The $WW$--fusion cross section has been studied in the
$\nu\bar{\nu}b\bar{b}$ final state for $M_H \le 160$~GeV. The contributions
to this final state from $WW$-fusion, Higgs--strahlung with 
$Z^0\to\nu\bar{\nu}$ and the remaining four--fermion final states can be 
separated, exploiting their different characteristics in the spectrum of the 
$\nu\bar{\nu}$ invariant mass, which is measurable through the missing mass 
distribution. From
a simultaneous fit to these contributions the $WW$--fusion cross section
can be extracted with accuracies between 2.8\% and 13\% for Higgs boson 
masses between 120 and 160~GeV~\cite{xs-hww}. The different behaviour of the
contributions to the missing mass spectrum for different polarisations of
the $e^+$ and the $e^-$ beam is advantageous to control systematic
uncertainties. The possibility to exploit the $\nu\bar{\nu}WW^*$ and
$\nu\bar{\nu}ZZ^*$ final states at higher values of $M_H$ is promising
to extend the mass reach of this measurement but has not yet been studied
in detail. 

The measurement of the branching ratio $H\to WW^*$ provides an alternative
means to access the $g_{HWW}$ coupling. This branching ratio can be measured in
Higgs--strahlung events and has been studied in the 
$Z\to\ell^+\ell^-,H\to q\bar{q}'q\bar{q}'$ and 
$Z\to q\bar{q}, H\to q\bar{q}'\ell\nu$ channels~\cite{br-hww}.
The analyses take advantage of the excellent jet-jet invariant mass
resolution achievable for future LC detectors under study. The achievable
precision is 5.1\% to 2.1\% for Higgs boson masses between 120 and 160~GeV. 

\begin{figure}[h!]
\begin{center}
\epsfig{file=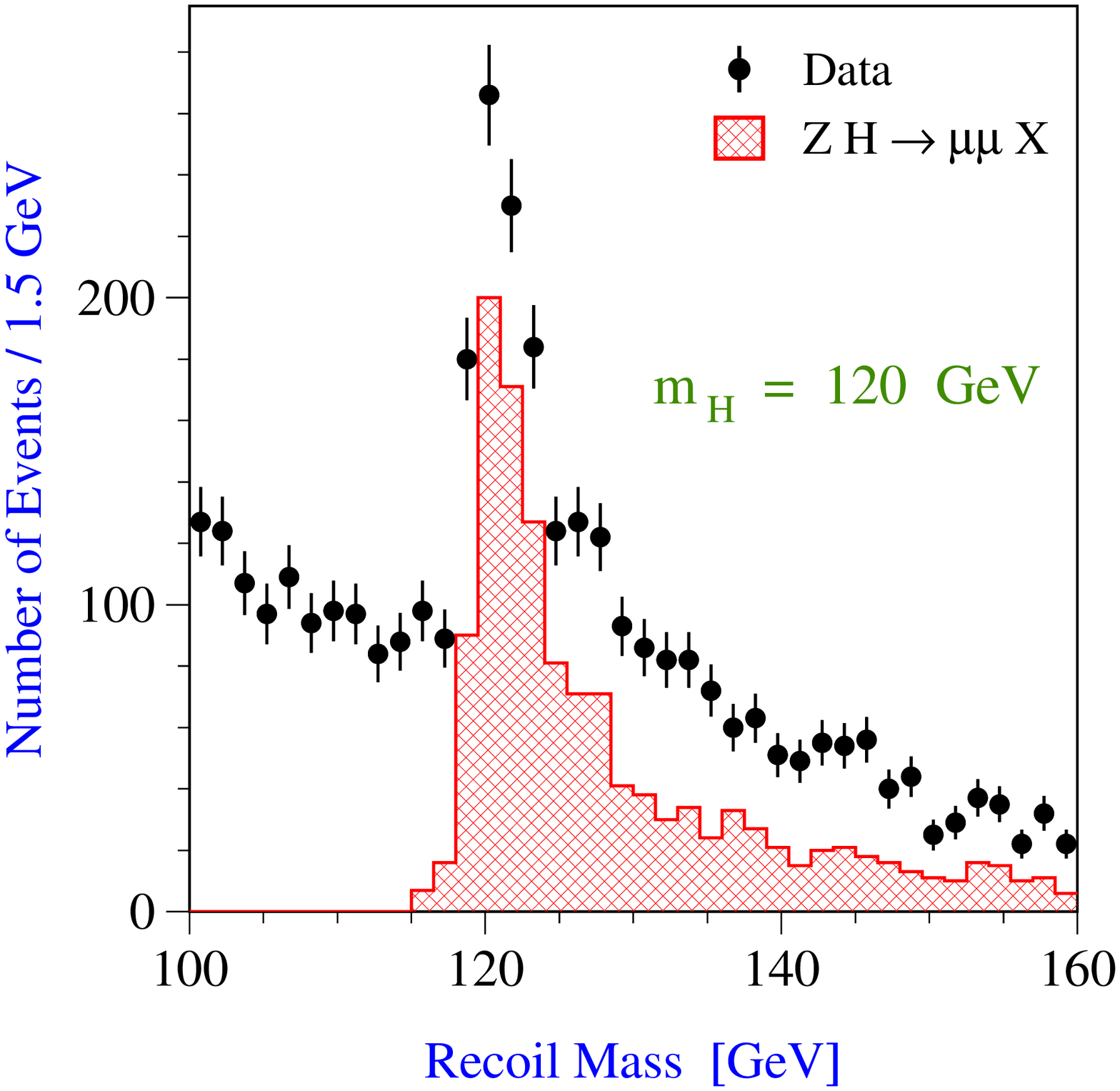,width=7.25cm,height=7.5cm,clip}
\epsfig{file=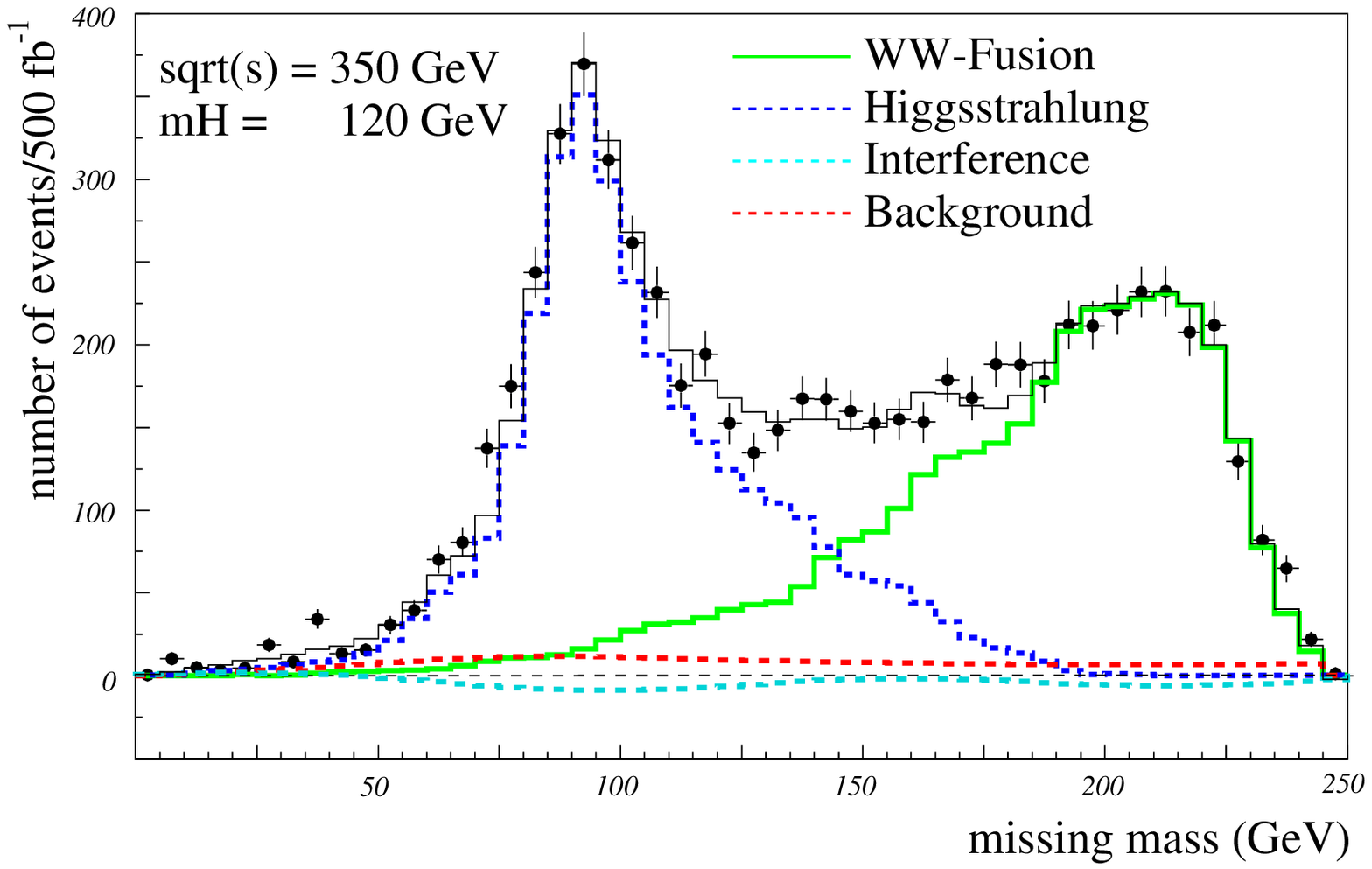,width=7.25cm,height=7.5cm,clip}
\caption{Observables for the Higgs couplings to the $Z^0$ and $W^{\pm}$ gauge
bosons. Left: di-lepton recoil mass in the Higgs-strahlung process $e^+e^- \to 
HZ \to X \ell^+\ell^-$. Right: missing mass distribution in 
$b\bar{b}\nu\bar{\nu}$ events. The $WW$~fusion contribution to the Higgs 
production is measured from a fit to the shape of this distribution.}
\label{fig:wwfusion}
\end{center}
\end{figure}

\begin{table}
\begin{center}
\caption{Relative accuracy on the determination of the Higgs production 
cross sections and decay branching ratios in gauge bosons for 500~fb$^{-1}$ of 
LC data at $\sqrt{s}$ = 350 or 500~GeV.}
\begin{tabular}{|l|c|c|c|}
\hline
        & $\delta \sigma/\sigma$ & $\delta \sigma/\sigma$ & 
$\delta \sigma/\sigma$ \\ 
        & $M_H$=120~GeV & $M_H$=140~GeV & $M_H$=160~GeV \\
\hline \hline
$\sigma(e^+e^- \to HZ)$ & $\pm$0.024 & $\pm$0.027 & $\pm$0.030 \\
$\sigma(e^+e^- \to H\nu_e \bar \nu_e)$ & $\pm$0.028 & $\pm$0.037 & 
$\pm$0.130 \\
BR($H \to WW^*$) & $\pm$0.051 & $\pm$0.025 & $\pm$0.021 \\ 
\end{tabular}
\end{center}
\end{table}

\subsection{Higgs Coupling to Photons}

The Higgs effective coupling to photons is mediated by loops, dominated in the 
SM by the $W$ contribution but also sensitive to any massive charged particles
coupling directly to the Higgs and to the photons. In the case of enhanced 
$Hbb$ or $Htt$ couplings or contributions from charged Higgs bosons, the 
$\gamma \gamma$ effective couplings may deviate significantly from its SM
prediction and provides insight into the structure of the Higgs 
sector~\cite{mariak}. This coupling can be tested both
through the $\gamma \gamma \rightarrow H$ production at a $\gamma \gamma$ 
collider and the Higgs decay channel $H \rightarrow \gamma \gamma$. The 
$\gamma \gamma \rightarrow H$ cross section being very substantial, a light
Higgs can be observed through its $b \bar b$ decay provided an effective 
suppression of the large $\gamma \gamma \rightarrow c \bar c$ background is
achieved. With an integrated luminosity of 150~fb$^{-1}$, an accuracy of 
2\% on $\sigma(\gamma \gamma \rightarrow H)$ can be achieved for a 120~GeV 
SM-like Higgs~\cite{hgamma}. The corresponding decay branching fraction can be
measured in both the $\nu \bar \nu \gamma\gamma$ and $\gamma\gamma+
{\mathrm{jets}}$ final states. The $e^+e^- \to Z\gamma\gamma$ double 
bremsstrahlung process represents the most important background process. This 
can be reduced by exploiting the photon energy and angular distributions.
Since the SM prediction for BR($H \to \gamma\gamma$) is only 
$2 \times 10^{-3}$, it can only be measured with an accuracy of 19\% for
500~fb$^{-1}$ of data. The error is reduced to 13\% for 
1000~fb$^{-1}$~\cite{reid1}.

\subsection{Extraction of Higgs Couplings}
The Higgs boson production and decay rates discussed above, can be used to 
measure 
the Higgs couplings to gauge bosons and fermions. Since some of the couplings
of interest can be determined independently by different observables while 
other determinations are partially correlated, it is interesting to perform
a global fit to the measurable observables to extract the Higgs couplings.
This method makes optimal use of the available information and can take 
properly into account the correlation originating from the experimental 
techniques.

\begin{table}[h!]
\begin{center}
\caption{Relative accuracy on Higgs couplings for 500~fb$^{-1}$ of LC data}
\begin{tabular}{|l|c|}
\hline
Coupling & $M_H$ = 120~GeV \\
\hline \hline
$g_{HWW}$ & $\pm$ 0.012    \\
$g_{HZZ}$ & $\pm$ 0.012    \\ \hline
$g_{Htt}$ & $\pm$ 0.022    \\
$g_{Hbb}$ & $\pm$ 0.021    \\
$g_{Hcc}$ & $\pm$ 0.031    \\ \hline
$g_{H\tau\tau}$ & $\pm$ 0.032  \\ \hline
\end{tabular}
\end{center}
\label{tab:hfitter}
\end{table}

\begin{figure}[ht!]
\begin{center}
\epsfig{file=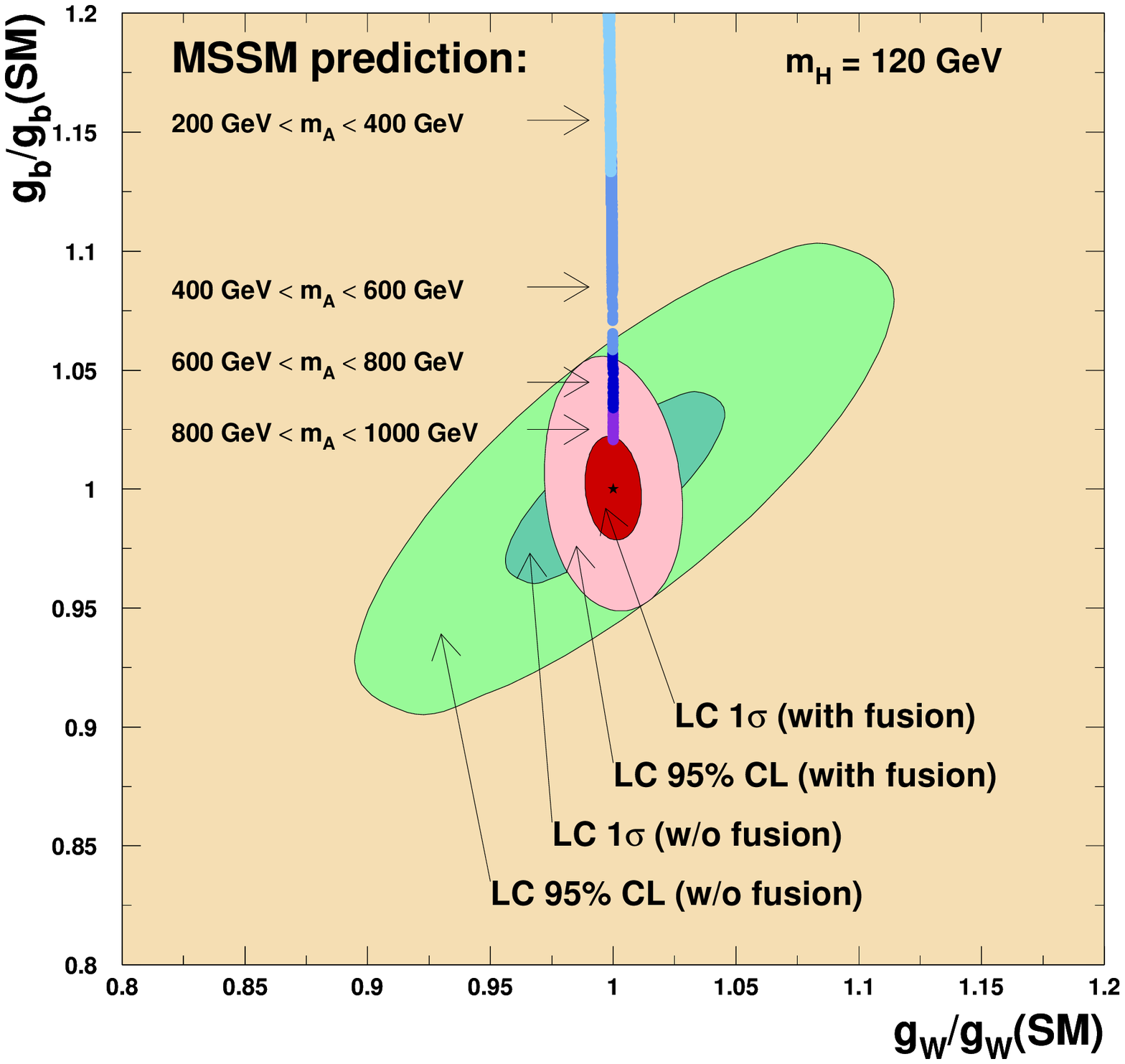,width=7.0cm,height=6.0cm,clip=}
\epsfig{file=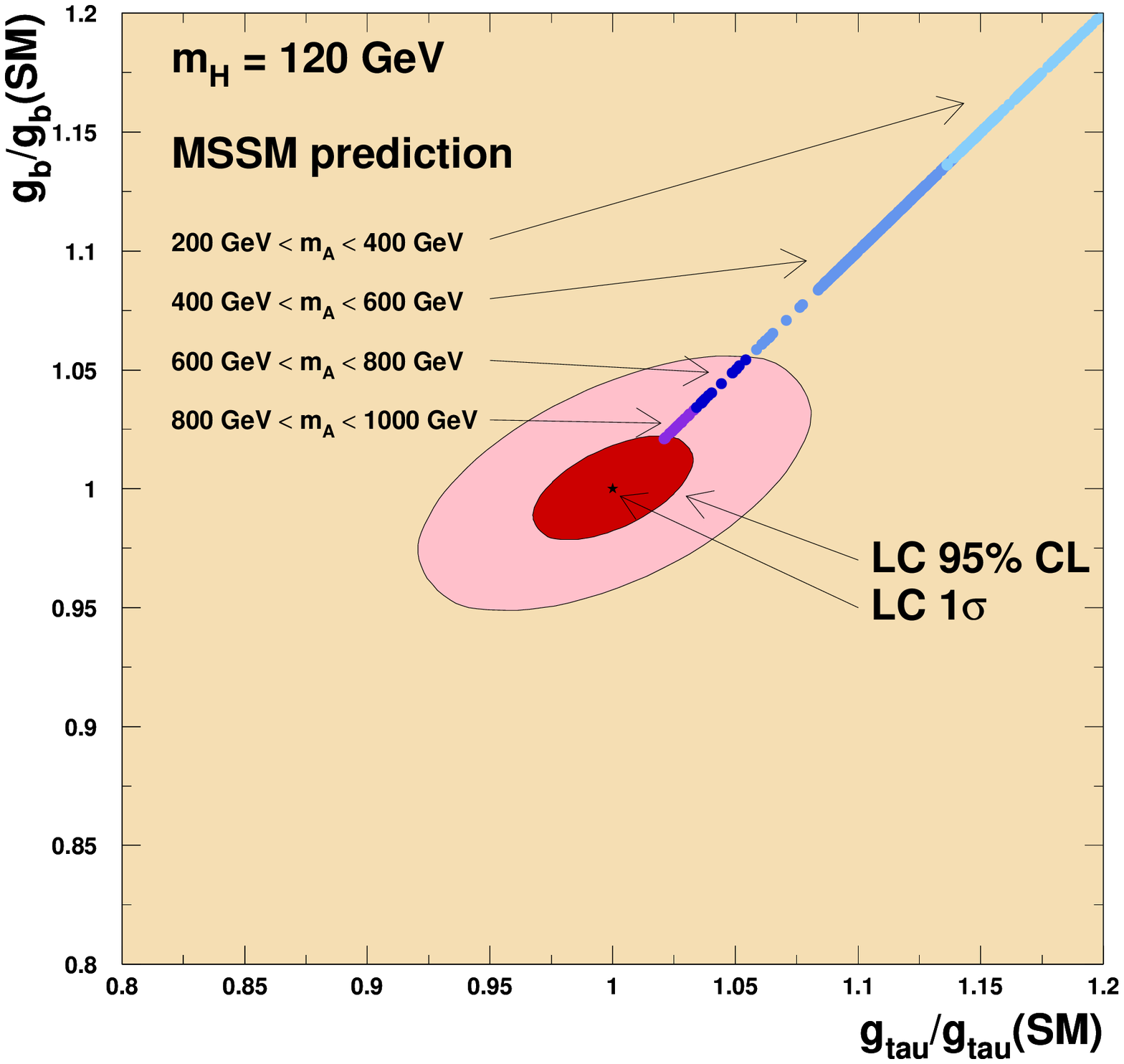,width=7.0cm,height=6.0cm,clip=}
\caption{Higgs coupling determinations at the LC. The contours for the 
$g_{Hbb}$ vs. $g_{HWW}$ (left) and $g_{Hbb}$ vs. $g_{H\tau\tau}$ (left) 
couplings for a 120~GeV Higgs boson as measured with 500~fb$^{-1}$ of data.}
\end{center}
\end{figure}
A dedicated program, {\sc HFitter}~\cite{hfitter} has been developed based on 
the {\sc Hdecay}~\cite{hdecay} program for the calculation of the Higgs boson
branching ratios. The following inputs have been used: $\sigma_{HZ}$, 
$\sigma_{H\nu\bar\nu}$, BR($H \rightarrow WW$), BR($H \rightarrow \gamma 
\gamma$), BR($H \rightarrow b \bar b$), BR($H \rightarrow \tau^+ \tau^-$),
BR($H \rightarrow c \bar c$), BR($H \rightarrow g g$), $\sigma_{t \bar t H}$.
For correlated measurements the full covariance matrix has been used.
The results are given for $M_H$ = 120~GeV and 500~fb$^{-1}$. 
Table~3 shows the accuracy which can be achieved in determining the couplings.

\subsection{Higgs Boson Width}

The total decay width of the Higgs boson is predicted to be too narrow
to be resolved experimentally for Higgs boson masses below the $ZZ$--threshold.
Above approximately 200 GeV the total width can be measured directly from
the reconstructed width of the recoil mass peak. 

For lower masses, indirect methods, exploiting relations between the total
decay width and the partial widths for exclusive final states, must be applied.
In general, the total width is given by 
$\Gamma_{tot}=\Gamma_X/\mathrm{BR}(H\to X)$. Thus whenever $\Gamma_X$ can be 
determined independently of the corresponding branching ratio, a measurement 
of $\Gamma_{tot}$ can be carried out. Two feasible options exist for light 
Higgs bosons: i) the extraction of $\Gamma_{WW}$ from the measurement of the 
$WW$--fusion cross section combined with the measurement of 
$\mathrm{BR}(H\to WW^*)$ and ii) the measurement of the $\gamma\gamma\to H$ 
cross section at a $\gamma\gamma$ collider combined with the measurement of 
$\mathrm{BR}(H\to \gamma\gamma)$ in $e^+e^-$ collisions.
The $WW$--fusion option yields a precision of 6\% to 13\% for
Higgs boson masses between 120 and 160 GeV, while the $\gamma\gamma$ option
yields a larger error dominated by the large uncertainty in the 
$\mathrm{BR}(H\to \gamma\gamma)$ determination discussed above.

Assuming the $SU(2)\times U(1)$ relation $g_{HWW}^2/g_{HZZ}^2 =
1/\cos^2{\theta_W}$ to be valid, the measurement of the Higgs--strahlung
cross section provides a viable alternative with potentially higher
mass reach than the $WW$--fusion option.

\subsection{Higgs Potential}

The observation of the scaling of the Higgs couplings to fermions with their
mass will provide with a proof that the interaction with the Higgs field
is responsible for the mass generation. However, in order to fully establish 
the Higgs mechanism, the Higgs potential 
$V = \lambda~(|\phi|^2-\frac{1}{2}v^2)^2$ with $v = (\sqrt{2}G_F)^{-1/2} 
\simeq$ 246~GeV must be reconstructed through the determination of the 
triple, $\lambda_{HHH}$, and quartic, $\lambda_{HHHH}$, Higgs self couplings. 
While effects from the quartic coupling may be too small to be observed at the
LC, the triple Higgs coupling can be measured in the double Higgs boson 
production processes $e^+e^- \rightarrow HHZ$ and $\nu_e\bar\nu_eHH$. In 
$e^+e^-$ collisions up to 1~TeV the double Higgs boson associated production 
with the $Z$ is favoured, while at a multi-TeV collider the $\nu_e\bar\nu_eHH$
reaction becomes dominant~\cite{vhiggs}. 
\begin{figure}[h!]
\vspace*{-0.25cm}
\begin{center}
\epsfig{figure=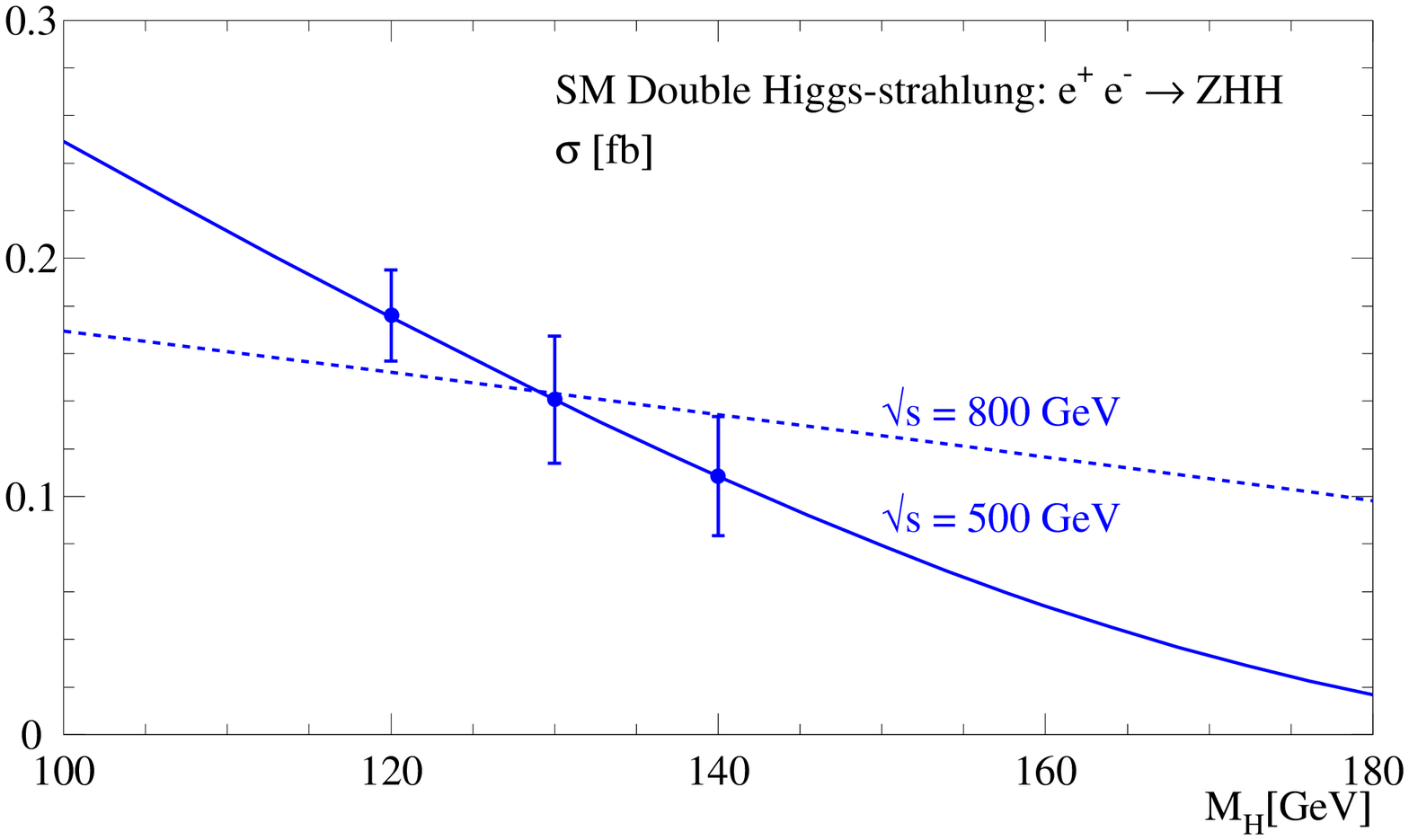,width=6.5cm,height=5.0cm}
\epsfig{figure=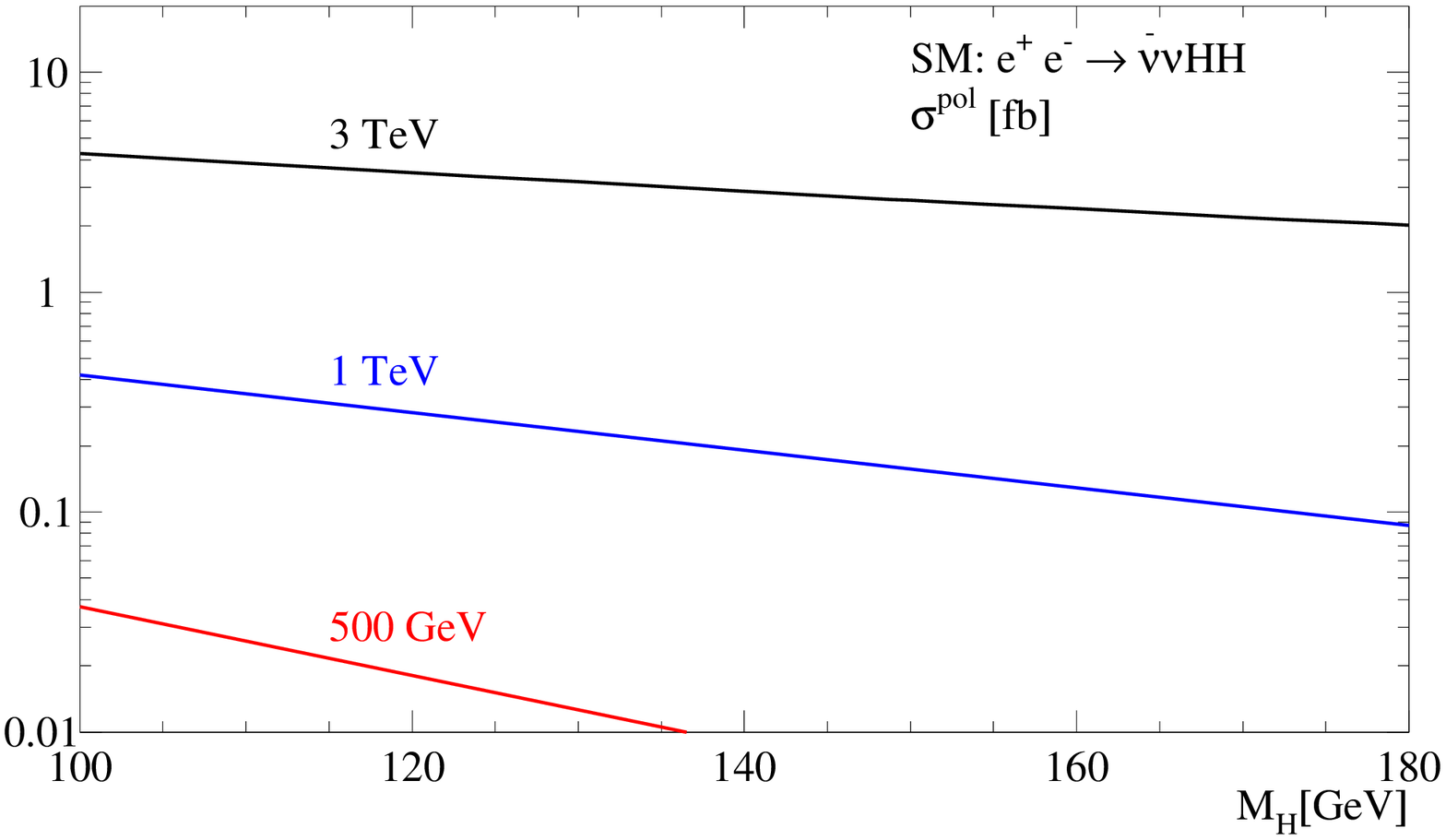,width=6.5cm,height=5.0cm}
\end{center}
\caption{Cross sections sensitive to the triple Higgs coupling.
$\sigma(e^+e^- \rightarrow HHZ)$ (left) and $\sigma(HH \nu \bar \nu)$ (right)
as a function of $M_H$ for different LC centre-of-mass energies.}
\label{fig:hhxs}
\vspace*{-0.25cm}
\end{figure}
The sensitivity to $\lambda_{HHH}$ from the measurement
of $\sigma_{HHZ}$ and $\sigma_{\nu\bar\nu HH}$ is diluted by the effects of 
other diagrams, not involving the triple Higgs coupling but leading
to the same final state. The four and six fermion backgrounds and the small
signal cross section make this measurement an experimental challenge . 
The distinctive 4 $b$-jet signature and the $M_{b \bar b} = M_H$ constraints 
allow to reduce these backgrounds to get $S/\sqrt{B} \simeq 6$ with 
1000~fb$^{-1}$ of data collected at $\sqrt{s}$ = 500~GeV, provided a
performant $b$-tagging and energy flow response of the detector~\cite{vexp}.
This corresponds to a determination of $\lambda_{HHH}$ with a statistical 
accuracy of 22\% for $M_H$ = 120~GeV with 1000~fb$^{-1}$. A second phase LC, 
delivering multi-TeV $e^+e^-$ collisions, could improve this accuracy to 
better than 10\%.

In the SM extensions with an extra Higgs doublet, additional trilinear Higgs 
couplings are also present such as $\lambda_{hhH}$, $\lambda_{hhA}$, 
$\lambda_{hhh}$ and $\lambda_{HAA}$. 
While these depend also on the $\tan \beta$ and $M_A$ 
parameters, the topologies analysed for the case of the SM also apply to
that of the $\lambda_{hhh}$ except for the limited region of parameters
where the $h \rightarrow b \bar b$ decay is suppressed. 
The corresponding analysis can be repeated for trilinear
Higgs couplings in the supersymmetric extension of the Standard Model.
The sensitive area in the $[M_A,\tan\beta]$ plane depends on the states that
can be analysed as described in detail in~\cite{triplesusy}.

\section{Higgs Bosons in Supersymmetry \label{sec:susy}}

Several extensions of the SM introduce additional Higgs doublets. In the 
simplest of such extensions (2HDM), the Higgs sector consists of two doublets 
generating five physical Higgs states: $h^0$, $H^0$, $A^0$ and $H^{\pm}$.
The $h^0$ and $H^0$ states are CP even and the $A$ is CP odd. 
Besides the masses, two mixing angles define the properties of the Higgs 
bosons and their interactions with gauge bosons and fermions defined through 
the ratio of the vacuum expectation values $v_2/v_1 = \tan \beta$ and a mixing
angle $\alpha$ in the neutral CP-even sector. Two Higgs doublets naturally 
arise in the context of the minimal supersymmetric extension of the SM (MSSM).
The study of the lightest neutral MSSM Higgs boson $h^0$ follows closely that
of the SM $H$ discussed above and those results remain in general valid. The 
ability of the LC to tell the SM/MSSM nature of a neutral Higgs, based solely 
on its properties, is discussed in the next section.
In SUSY models, additional decay channels may open for the $h^0$ boson, if 
there are light enough SUSY particles. 
The most interesting scenario is that in which the Higgs decays in 
particles escaping detection, such as $h^0 \rightarrow \chi^0 \chi^0$, giving 
a sizeable $H \rightarrow {\mathrm{invisible}}$ decay width. While the 
Higgs observability in the dilepton recoil mass in the associated $HZ$ 
production channel is virtually unaffected by this scenario, such an invisible
decay width can be measured by comparing the number of 
$e^+e^- \rightarrow ZH \rightarrow \ell^+\ell^- {\mathrm{anything}}$ events 
with the sum over the visible decay modes corrected by the $Z^0 \rightarrow
\ell^+ \ell^-$ branching fraction: BR($Z \rightarrow \ell^+\ell^-$) $\times
(\sum_{i=b,c,\tau,...} N_{ZH \rightarrow f_i \bar f_i} + 
\sum_{j=W,Z,\gamma} N_{ZH \rightarrow B_j \bar B_j}$). By taking the accuracies
on the determination of the individual branching fractions discussed above,
the rate for the $H \rightarrow {\mathrm{invisible}}$ decay can be determined 
to better than 20\% for BR($H \rightarrow {\mathrm{invisible}}$)$>$0.05.

\subsection{Tell the SM from a MSSM Neutral Higgs}

The discovery of a neutral Higgs boson, with mass in the range 115~GeV $< M_H 
<$ 140~GeV, will raise the question of whether the observed particle is the SM
Higgs boson or the lightest boson from the Higgs sector of a SM extension.
It has been shown that, for a large fraction of the $\tan \beta - M_A$ 
parameter plane in the MSSM, this neutral boson will be the only Higgs state
observed at the LHC (see Figure~10). It is possible, that the scale $M_{SUSY}$
is high and
the supersymmetric fermion partners may not be visible at a 500~GeV linear
collider. In this circumstance, a Higgs particle generated by a complex 
multi-doublet model could be indirectly recognised only by a study of its 
couplings. If the $HZZ$ coupling, measured by the Higgs-strahlung production 
cross-section independently from the Higgs boson decay mode, turns out to be 
significantly smaller than the SM expectation, this will signal the existence 
of extra Higgs doublets.

The determination of the Higgs boson decay branching ratios with the accuracy
anticipated by these studies can be employed to identify the SM or MSSM 
nature of a light neutral Higgs boson. 
The Higgs boson decay widths $\Gamma^{MSSM}$ to a specific final state
are modified as follows with respect to the SM $\Gamma^{SM}$:
$\Gamma^{MSSM}_{b \bar b} \propto \Gamma^{SM}_{b \bar b} 
\frac{\sin^2 \alpha}{\cos^2 \beta}$ and
$\Gamma^{MSSM}_{c \bar c} \propto \Gamma^{SM}_{c \bar c} 
\frac{\cos^2 \alpha}{\sin^2 \beta}$.
Therefore, deviations in the ratios of branching ratios such as
{$\frac{BR(h \rightarrow W W^*)}{BR(h \rightarrow b \bar b)}$,
$\frac{BR(h \rightarrow c \bar c)}{BR(h \rightarrow b \bar b)}$ and
$\frac{BR(h \rightarrow g g)}{BR(h \rightarrow b \bar b)}$ from their SM 
expectations can reveal the MSSM nature of the Higgs boson and also provide 
indirect information on the mass of the $CP$-odd $A^0$ Higgs boson, even when
it is so heavy that it can not be directly observed at $\sqrt{s}$ = 500~GeV. 

To compare the SM predictions with those in MSSM, a complete scan of the MSSM 
parameter space has been performed.
For each set of parameters, the $h^0$ mass has been computed using the
diagrammatic two-loop result~\cite{fhiggs1}. Solutions corresponding to 
$M_{h^0} = (120 \pm 2)$~GeV have been selected and used to compute the 
$h^0$ decay branching ratios accounting for squark loops~\cite{hdecay}. 
The deviations from the SM predictions for
BR($h \rightarrow b \bar b$)/BR($h \rightarrow$ hadrons),
BR($h \rightarrow c \bar c$)/BR($h \rightarrow$ hadrons),
BR($h \rightarrow g g$)/BR($h \rightarrow$ hadrons) and
BR($h \rightarrow b \bar b$)/BR($h \rightarrow W W^*$) have been used to 
investigate the SM/MSSM discrimination. Figure~8 shows the region of the 
$M_A - \tan \beta$ plane in which there are more than 68\%, 90\% or 95\% of 
the MSSM solutions outside the SM 95\% confidence level region defined by the 
total $\chi^2$ probability for the observed branching ratios. 

\begin{figure}[htb]
\begin{center}
\epsfig{file=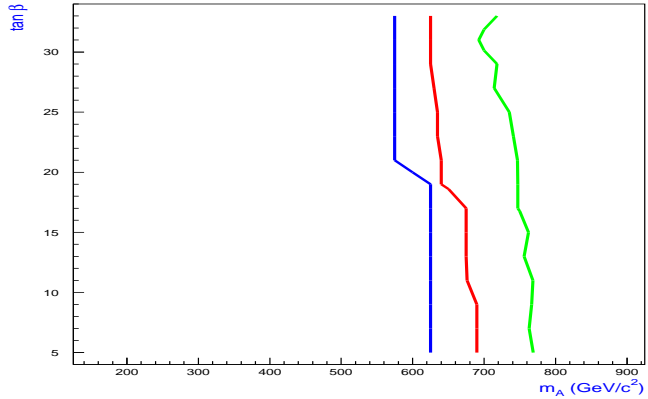,width=9.5cm,height=6.0cm,clip}
\caption{Sensitivity to SM/MSSM Higgs boson nature as a function of 
$M_A$ and $\tan \beta$ from determinations of $h^0$ BRs at the LC with 
1000~fb$^{-1}$. The curves show the 
upper bound of the regions with 68\%, 90\% and 95\% (from right to left) 
of the MSSM solutions distinguishable from the SM branching fraction 
predictions.}
\end{center}
\label{fig:matgb}
\end{figure}             

If a significant indication of MSSM has been observed, which implies that
$M_A$ is within the limit of Figure~8, then it is possible to go
further and use the accurate measurements of the Higgs boson decays to obtain 
an indirect estimate of the mass $M_{A^0}$. The sensitivity
to the $A^0$ mass arises from the MSSM corrections to the Higgs couplings
discussed above and it is of special interests for those masses above the
kinematical limit for direct $e^+e^- \rightarrow h^0 A^0$, $H^0 A^0$ 
production. 
The analysis has been performed assuming given sets of measured values for the 
BR($h \rightarrow c \bar c$)/BR($h \rightarrow b \bar b$),
BR($h \rightarrow g g$)/BR($h \rightarrow b \bar b$) and 
BR($h \rightarrow W W^*$)/BR($h \rightarrow b \bar b$) ratios. The $A^0$ mass 
has been varied together with the other MSSM parameters within the range 
compatible with the measured branching ratios allowing for their total 
uncertainty. The range of values of $M_A$ for the accepted MSSM solutions 
corresponds to an accuracy of 70~GeV to 100~GeV for the indirect
determination of $M_A$ in the mass range 300~GeV $< M_A <$ 600~GeV.

\subsection{Properties of the Heavy Higgs Sector}

A most distinctive feature of extended models such as supersymmetry, or general
2HDM extensions of the SM, is the existence of additional Higgs bosons.
Their mass and coupling patterns vary with the model parameters. However in
the decoupling limit, the $H^{\pm}$, $H^0$ and $A^0$ bosons are expected to be
heavy and to decay predominantly into quarks of the third generation.
\begin{figure}[h!]
\begin{center}
\epsfig{file=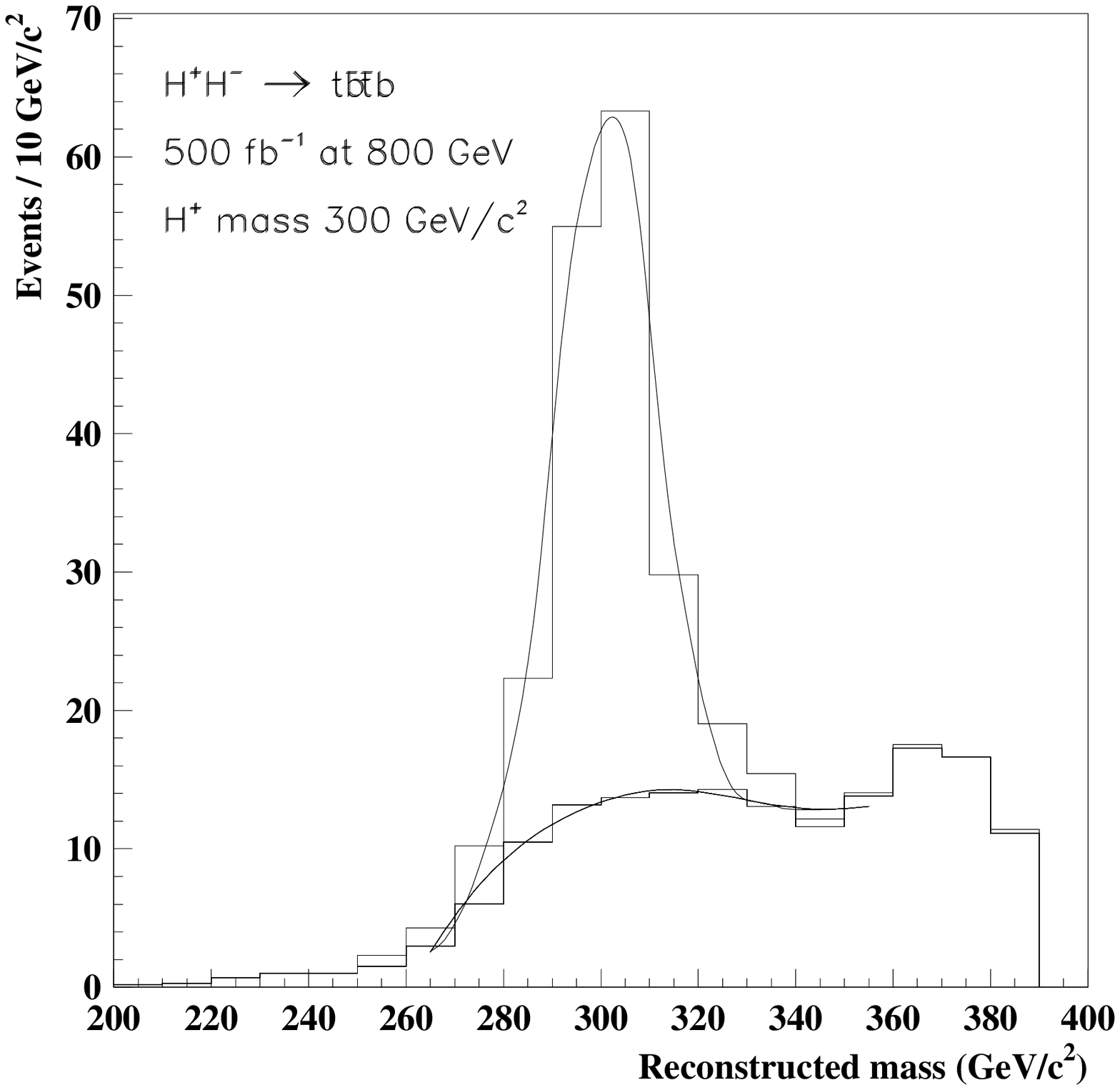,bbllx=20,bblly=140,bburx=535,bbury=675,height=6.0cm,
width=6.5cm}
\epsfig{file=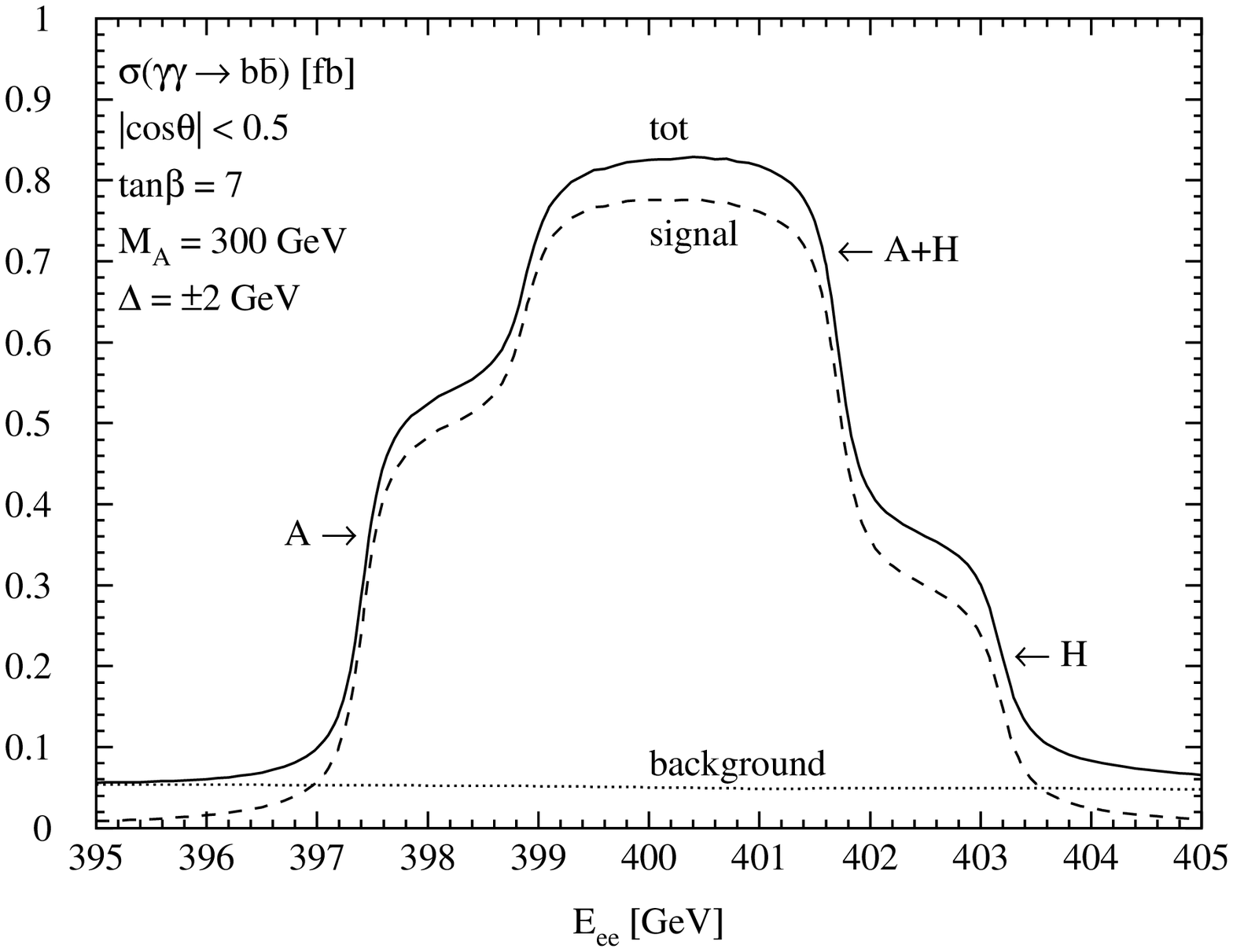,bbllx=35,bblly=225,bburx=575,bbury=635,height=6.0cm,
width=7.0cm} 
\caption{Examples of the signals from 300~GeV SUSY Higgs bosons at the LC for
$e^+e^-$ and $\gamma \gamma$ collisions.
Left: Di-jet invariant mass distribution for $e^+ e^- \rightarrow 
H^+H^- \rightarrow t \bar b \bar t b$ charged Higgs boson events after 
intermediate $W$ and $t$ mass and equal mass final state constraints.
Right: Scan of the $\gamma \gamma \to A^0$ and $H^0$ thresholds showing the 
sensitivity to the small mass splitting of nearly degenerate $A^0$ and $H^0$ 
Higgs bosons.}
\end{center}
\label{fig:hagg}
\end{figure} 

Establishing their existence and the determination of their mass and
main decay modes, through their pair production $e^+e^- \to H^0 A^0$, $H^+H^-$
will represent an important part of the LC physics programme at 
$\sqrt{s} \ge$~500~GeV. The decay channels $H^0$, $A^0 \to b \bar b$, 
$H^+ \rightarrow t \bar b$ or $W^+ h^0$, $h^0 \to b \bar b$ will provide with 
very distinctive 4~jet and 8~jet final states with 4~$b$-quark jets that can 
be efficiently identified and reconstructed. Exemplificative analyses have been
performed for these channels, showing that an accuracy of about 0.3\% on their
mass and of $\simeq 10\%$ on $\sigma \times BR$ can be obtained~\cite{hpm,ha}. 
In addition, $\gamma \gamma \to H^+H^-$  pair production and $\gamma \gamma
\to A^0$ and $H^0$ at the $\gamma \gamma$ collider, is characterised by
sizable cross sections and may also probe the heavier part of the Higgs 
spectrum in SM extensions. In particular, a scan of the $A^0$ and $H^0$ 
thresholds at the $\gamma \gamma$ collider can resolve a small $A^0 - H^0$ 
mass splitting~\cite{ggha} (see Figure~9). A detailed analysis of 
their production and decay, using polarised colliding photons, provides an 
opportunity to determine the quantum numbers of the heavy bosons and therefore
to distinguish the $A^0$ from the $H^0$ boson~\cite{ggha2}.  

\section{The LC, the LHC and the FMC \label{sec:comp}}

While the LC offers an accurate and comprehensive test of the Higgs 
sector properties, the LHC $pp$ collider, under construction at CERN, 
will independently probe the Higgs sector of the SM and its extensions and 
a muon collider, through the s-channel Higgs production, also has a claim
to the precision study of a light Higgs boson. We summarise here the main 
arguments on the complementary role of the LC compared to the Higgs boson 
picture that may be obtained on the basis of the LHC data, as anticipated by
the present studies, and compare the LC reach to that of a muon collider.

\begin{figure}[h!]
\begin{center}
\vspace*{-0.75cm}
\epsfig{file=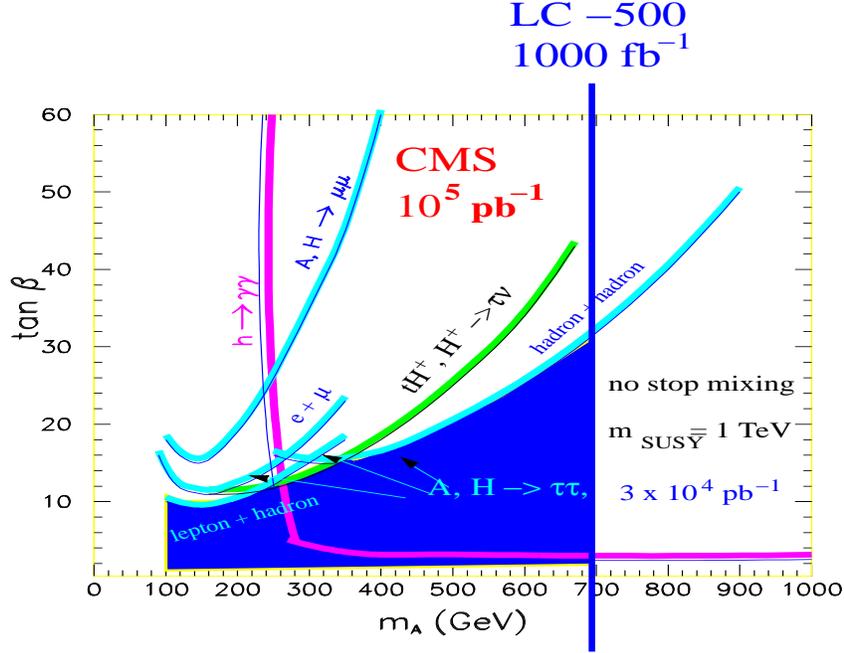,width=11.5cm,height=9.0cm,clip=}
\caption{Comparison of the MSSM Higgs sector discovery potential at the 
LHC and at the LC. The LHC data, from the CMS experiment simulation, show the
portions of the $M_A$ - $\tan \beta$ plane where the extended nature of the 
Higgs sector can be established by the observation of additional Higgs bosons. 
The accurate measurement of the lightest Higgs boson couplings at the LC for 
$\sqrt{s}$ = 500~GeV allows to complement this information by revealing the 
Higgs boson nature in the dark shaded region, for $M_A$ up to $\simeq$~700~GeV.}
\end{center}
\label{fig:cms-lc}
\end{figure}
At the LHC the SM Higgs boson, or at least one Higgs boson in SUSY extensions,
will be observed, unless it decays invisibly, as discussed above, while 
scenarios with extended Higgs sectors or non standard Higgs decay modes have 
been proposed that may evade the LHC probe of the existence of a Higgs boson. 
\begin{table}[h!]
\begin{center}
\caption{Summary of the Higgs boson profile from the LHC and the LC data.
Relative accuracies for the measurement of Higgs properties for different
Higgs boson masses.}
\begin{tabular}{|l|c|c|c|}
\hline 
        & $M_H$            & $\delta(X)/X$ & $\delta(X)/X$ \\
        & (GeV/$c^2$)      & LHC & LC \\
        &                  & 2 $\times$ 300 fb$^{-1}$ & 500 fb$^{-1}$ \\ 
\hline \hline
$M_H$      & 120           & ~9 $\times 10^{-4}$ & 3 $\times 10^{-4}$ \\
$M_H$      & 160           & 10 $\times 10^{-4}$ & 4 $\times 10^{-4}$ \\ 
\hline
$\frac{g_{Htt}}{g_{HWW}}$ & 120 & 0.070 & 0.023  \\
$\frac{g_{HZZ}}{g_{HWW}}$ & 160 & 0.050 & 0.022  \\ \hline
$\Gamma_{tot}$ & 120-140 &   -     & 0.04 - 0.06 \\ \hline
$g_{Huu}$ & 120-140 & - & 0.02 - 0.04 \\
$g_{Hdd}$ & 120-140 & - & 0.01 - 0.02 \\
$g_{HWW}$ & 120-140 & - & 0.01 - 0.03 \\ \hline
$\Phi = H + A$ & 120-140 & - &  0.03 - 0.13 \\
$\lambda_{HHH}$ & 120 & - & 0.22 \\ \hline 
\end{tabular}
\end{center}
\label{tab:lhc1}
\end{table}
Beyond discovery, measuring the Higgs properties in $pp$ collisions is 
difficult due to the limited signal statistics, large backgrounds and 
systematic uncertainties arising, for example from the limited precision of the
parton densities in the proton. Still, the LHC data can provide ratios of some
of the Higgs branching ratios and couplings, as listed in 
Table~4 for $M_H = 120$ GeV, while the precision measurements of the absolute 
branching ratios and couplings remain an experimental program which can only be
addressed at a high luminosity lepton collider. 

\begin{figure}[h!]
\begin{center}
\vspace*{-0.75cm}
\epsfig{file=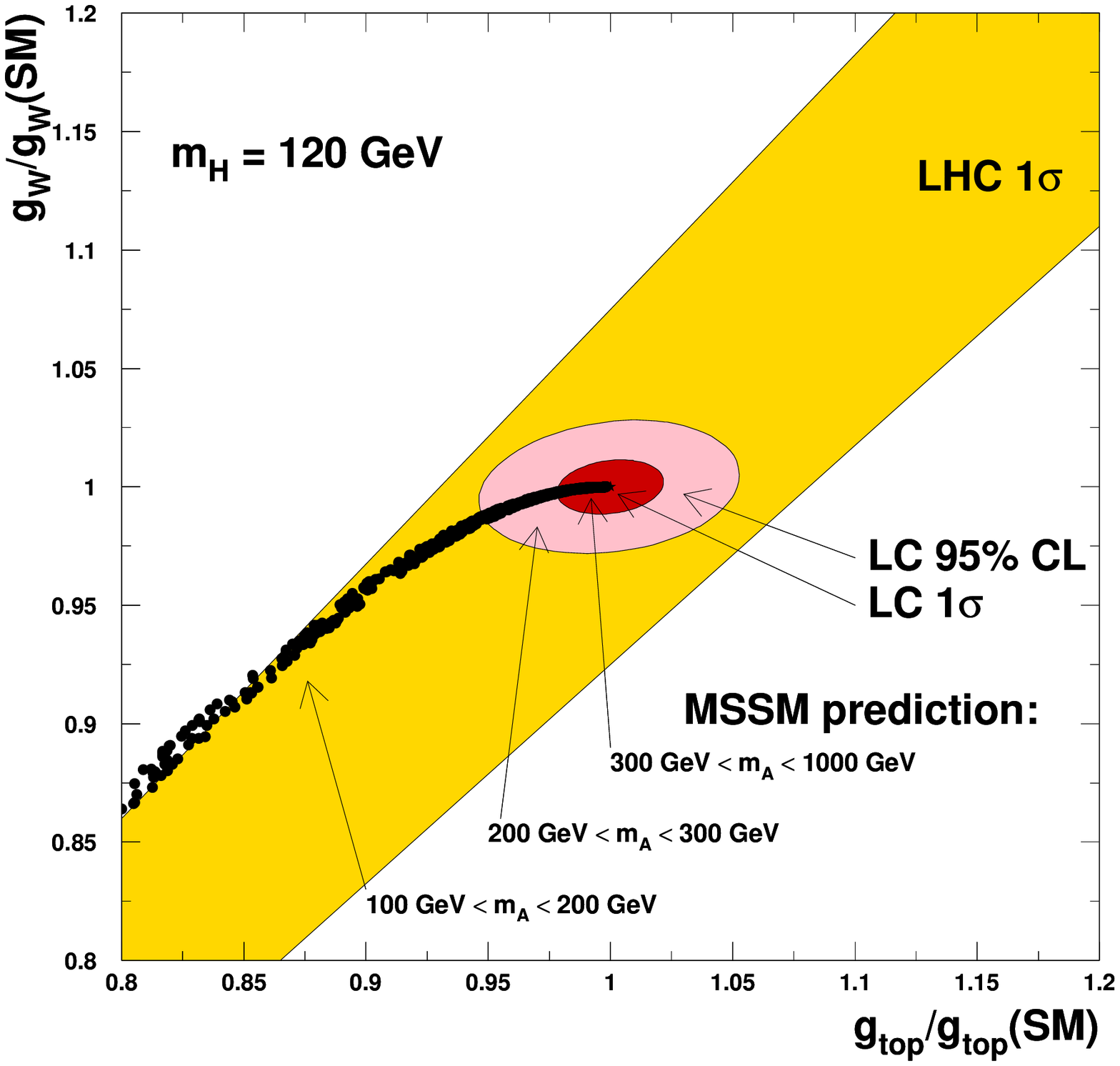,width=10.0cm,height=8.5cm}
\caption{Accuracy in the determination of the $g_{Htt}$ and $g_{HWW}$ 
Higgs boson couplings at the LHC and at the LC compared to the 
predictions from MSSM for different values of the $M_A$ mass.}
\end{center}
\label{fig:lhclcgh}
\end{figure}
Further perspectives for measurements of the total Higgs 
boson width and of the $g_{HWW}$ coupling with a precision of 10--20\% have 
been recently suggested for the LHC~\cite{zeppenfeld} but are still awaiting 
to be confirmed experimentally, accounting for the backgrounds and the 
detector response. Since over a large range of the parameter space, only one 
Higgs boson is observable at the LHC, the distinction of the SM or SUSY nature 
of the Higgs sector can only be addressed at a lepton collider, while the 
LHC may discover supersymmetry by directly observing sparticle decays.

The complementarity of the linear collider data to the picture of the Higgs 
sector as it will have been outlined by the LHC is therefore threefold: 
i) the accuracy of those measurements, which are possible at the LHC, can be 
significantly increased, ii) the absolute measurements of all the relevant 
Higgs boson couplings, including the Higgs self coupling, will be possible 
only at the lepton collider and iii) extended Higgs sector scenarios (e.g. 
invisible Higgs boson decays or 2HDM) can be observed at the linear collider 
closing the loopholes of a possible non-discovery at the LHC.

If the considerable technical challenges presented by a muon collider project 
can be properly addressed, a first muon collider (FMC)~\cite{fmc,cernfmc} 
could be operated at the centre-of-mass energy corresponding to a light Higgs 
boson mass. Such a machine would provide with the unique feature of s-channel 
$\mu^+\mu^- \rightarrow H$ production with a beam energy spread comparable to 
the natural width of a light SM Higgs boson. However such a small energy 
spread may only be achieved at a luminosity of about 
$10^{31}$~cm$^{-1}$s$^{-1}$, corresponding to the production of a few thousand 
Higgs particles/year. 
By an energy scan of the Higgs resonance, the FMC can determine its mass and 
width to an accuracy of better than 1~MeV. The measurement of the 
$\mu^+\mu^- \rightarrow H \rightarrow b \bar b$ cross section would offer 
sensitivity to the Higgs couplings to the $b$ quark. This measurement can thus 
be used, together with those of the Higgs width and mass for a test the Higgs 
boson nature, similarly to that discussed above for the case of the $e^+e^-$
collider. A comparable sensitivity to the MSSM Higgs nature can be obtained 
with 200~pb$^{-1}$, while a significant extension towards larger values of 
$M_A$ would require an order of magnitude higher luminosity~\cite{cernfmc}. 
Considering the different maturity of the LC and FMC projects, the LC appears 
at present as the most attractive option for a detailed investigation of the 
Higgs sector, complementing the data from the Tevatron and the LHC, at the turn
of the first decade of this new century. 

\section{Summary}

The search for the Higgs boson and the study of its properties is one of the 
main goals of present research in particle physics. The central role of a
linear collider in the understanding of the mechanism of electro-weak symmetry 
breaking, complementing the data delivered by LEP and those to be acquired 
at the Tevatron and at the LHC, has been clearly outlined by the studies 
carried out world-wide, reported at this Conference and summarised in this 
review.    

\section*{Acknowledgements}

We are grateful to the organisers of the LCWS-2000 Conference for their
invitation and the inspiring atmosphere. The results reported here are due to 
the activities of many colleagues, in the regional Higgs working groups of the 
Worldwide Study on Physics at Linear Colliders, whose contributions are
gratefully acknowledged.


\begin{references}

\bibitem{Higgs} P.W.~Higgs, {\it Phys. Rev. Lett.} {\bf 12} (1964) 132;
{\it idem}, {\it Phys. Rev.}  {\bf 145} (1966) 1156; F.~Englert and R.~Brout, 
{\it Phys. Rev. Lett.} {\bf 13} (1964) 321; G.S.~Guralnik, C.R.~Hagen and 
T.W.~Kibble, {\it Phys. Rev. Lett.} {\bf 13} (1964) 585.

\bibitem{triv} A.~Hasenfratz {\it et al.}, {\it Phys. Lett.} {\bf B199} 
(1987) 531; M.~L\"uscher and P.~Weisz, {\it Phys. Lett.} {\bf B212} (1988) 472;
M. G\"ockeler {\it et al.}, {\it Nucl. Phys.} {\bf B404} (1993) 517.

\bibitem{pik} P.~Igo-Kemenes for the LEP Higgs working group, 
talk given at the open session of the LEPC comittee, CERN, Nov 2000;
see also: R.~Barate {\it et al.} (ALEPH Coll.), {\it Phys. Lett.} {\bf B495} 
(2000) 1; M.~Acciari {\it et al.} (L3 Coll.), {\it Phys. Lett.} {\bf B495} 
(2000) 18.

\bibitem{lepewwg}
The LEP Collaborations, {\it A Combination of Preliminary Electroweak 
Measurements and Constraints on the Standard Model}, CERN-EP Note 
in preparation.

\bibitem{bes}
Z.G.~Zhao (BES Coll.), to appear in the proceedings of the XXX$^{th}$ 
{\it Int. Conf. on High Energy Physics}, Osaka, July~2000 and hep-ex/0012038.

\bibitem{TevHiggs} M.~Carena {\it et. al.}, {\it Report of the Tevatron
Higgs working group}, hep-ph/0010338.

\bibitem{caner}
A.~Caner, to appear in the proceedings of the XXX$^{th}$ {\it Int. Conf.
on High Energy Physics}, Osaka, July~2000.

\bibitem{hmass1}
P.G.~Abia, these proceedings.

\bibitem{DK} 
A.~Djouadi and B.A.~Kniehl, DESY~93-123C, 51.

\bibitem{dmiller}
D.~Miller, these proceedings.

\bibitem{hqn1}
M.~Schumacher, LC-PHSM-2001-003.

\bibitem{HIKK} 
K.~Hagiwara {\it et al.}, {\it Eur.\ Phys.\ J.} {\bf C14} (2000) 457 and
J.~Kamoshita, these proceedings.

\bibitem{brhad0}
M.D.~Hildreth, T.L.~Barklow and D.L.~Burke, {\it Phys. Rev. Lett.} {\bf 49}
(1994), 3441.

\bibitem{brhad1}
M.~Battaglia, in Proc. of the {\it Worldwide Study on Physics 
and Experiments with Future $e^+e^-$ Linear Colliders}, E.~Fernandez and 
A.~Pacheco (editors), UAB, Barcelona 2000, vol.~I, 163 and hep-ph/9910271.

\bibitem{brhad2} 
J.~Brau, these proceedings.

\bibitem{BCDKZ} 
V.~Barger {\it et al.}, {\it Phys.\ Rev.}\ {\bf D49} (1994) 79; \\
K.~Hagiwara and M.L.~Stong, {\it Z. Phys.}\ {\bf C62} (1994) 99.

\bibitem{ttH} 
A.~Djouadi, J.~Kalinowski and
P.M.~Zerwas, {\it Mod. Phys. Lett.} {\bf A7} (1992) 1765 and 
{\it Z. Phys.} {\bf C54} (1992) 255.

\bibitem{ttHth}
S.~Dittmar {\it et al.}, {\it Phys. Lett.} {\bf B441} (1998), 383 and 
{\it Phys. Lett.} {\bf B478} (2000), 247; S.~Dawson and L.~Reina, {\it Phys.
Rev. Lett.} {\bf D57} (1998), 5851 and {\it Phys. Rev.} {\bf D59} (1999), 
054012 and S.~Dawson, these proceedings.

\bibitem{tthlight}
A.~Juste and G.~Merino, in Proc. of the {\it Worldwide Study on Physics 
and Experiments with Future $e^+e^-$ Linear Colliders}, E.~Fernandez and 
A.~Pacheco (editors), UAB, Barcelona 2000, vol.~I, 265 and hep-ph/9910301.

\bibitem{ttheavy}
J.~Alcarez and E.~Ruiz~Morales, these proceedings and hep-ph/0012109

\bibitem{xs-hww}
K.~Desch and N.~Meyer, these proceedings.

\bibitem{br-hww}
G.~Borisov and F.~Richard, Note LAL~99-26 and hep-ph/9905413.

\bibitem{mariak}
M.~ Krawczyk, these proceedings.

\bibitem{hgamma}
G.~Jikia and S.~S\"oldner-Rembold, {\it Nucl. Phys. Proc. Suppl.} {\bf 82} 
(2000) 373,

\bibitem{reid1}
E.~Boos {\it et al.}, LC-PHSM-2000-053 and 
H.J.~Schreiber {\it et al.}, these proceedings.

\bibitem{hfitter}
K.~Desch and M.~Battaglia, these proceedings.

\bibitem{hdecay}
A.~Djouadi, M.~Spira and P.~Zerwas, {\it Z. Phys.} {\bf 70} (1996) 427;\\
A.~Djouadi, J.~Kalinowski and M.~Spira, {\it Comput. Phys. Commun.} {\bf 108}
(1998), 56.

\bibitem{vhiggs}
A.~Djouadi {\it et al.}, {\it Eur. Phys. J.} {\bf C10} (1999) 27 and\\
M.~M\"uehlleitner, these proceedings.

\bibitem{vexp}
P.~Gay, these proceedings.

\bibitem{triplesusy} G. Gounaris, 
A.~Djouadi, H.E.~Haber and P.M.~Zerwas, {\it Phys. Lett.} {\bf B375} (1996) 
203; A.~Djouadi {\it et al.}, {\it Eur. Phys. J.} {\bf C10} (1999) 27 and 
M.~M\"uhlleitner, these proceedings.

\bibitem{fhiggs1}
S.~Heinemeyer, W.~Hollik and G.~Weiglein, {\it Phys. Lett.}{\bf B440} (1998), 
296 and {\it Phys. Rev.} {\bf D58} (1998) 091701;\\
S.~Heinemeyer, W.~Hollik and G.~Weiglein, {\it Comput. Phys. Commun.} 
{\bf 124} (2000) 76.

\bibitem{hpm}
A.~Kiiskinen, M.~Battaglia and P.~P\"oyh\"onen, these proceedings.

\bibitem{ha}
A.~Andreazza and C.~Troncon, DESY-123-E, 417.

\bibitem{ggha}
M.~M\"uehlleitner, these proceedings.

\bibitem{ggha2}
E.~Asakawa, these proceedings.

\bibitem{zeppenfeld} 
D.~Zeppenfeld {\it et al.}, {\it Phys. Rev.} {\bf D62} (2000) 13009.

\bibitem{fmc}
C.~Ankenbrandt {\it et al.}, {\it Phys. Rev. ST Accel. Beams} {\bf 2} (1999)
081001 (1999)
 
\bibitem{cernfmc}
{\it Prospective Study of Muon Storage Rings at CERN}, B.~Autin, A.~Blondel and
J.~Ellis (editors), CERN-99-02. 
 
\end{references}
\end{document}